# Using SHAP Values and Machine Learning to Understand Trends in the Transient Stability Limit

Robert I. Hamilton and Panagiotis N. Papadopoulos, *Member, IEEE*.

*Abstract*— **Machine learning (ML) for transient stability assessment has gained traction due to the significant increase in computational requirements as renewables connect to power systems. To achieve a high degree of accuracy; black-box ML models are often required – inhibiting interpretation of predictions and consequently reducing confidence in the use of such methods. This paper proposes the use of SHapley Additive exPlanations (SHAP) – a unifying interpretability framework based on Shapley values from cooperative game theory – to provide insights into ML models that are trained to predict critical clearing time (CCT). We use SHAP to obtain explanations of location-specific ML models trained to predict CCT at each busbar on the network. This can provide unique insights into power system variables influencing the entire stability boundary under increasing system complexity and uncertainty. Subsequently, the covariance between a variable of interest and the corresponding SHAP values from each location-specific ML model – can reveal how a change in that variable impacts the stability boundary throughout the network. Such insights can inform planning and/or operational decisions. The case study provided demonstrates the method using a highly accurate opaque ML algorithm in the IEEE 39-bus test network with Type IV wind generation.**

## I. Introduction

THE global decarbonization agenda is leading to the retirement of carbon intensive synchronous generation (SG) in favour of intermittent non-synchronous renewable energy resources. The complex highly non-linear dynamics associated with RES are proving challenging for maintaining system stability [1] as many of the attributes that SGs provide are lost/reduced (e.g., inertia). This is particularly notable in the transient stability problem [2]. In [3], the impact of RES is shown to have a widely varying impact on the transient stability boundary (both improvement and deterioration). Trends are often location-specific and depend on several parameters, which is often very complicated to understand.

Transient stability assessment (TSA) has typically been conducted via root-mean square (RMS) time-domain simulations (TDS) or transient energy function (TEF) methods, which are derived from Lyapunov stability theory [4]. While accurate, a key disadvantage of RMS TDS is relatively slow computational speed, particularly when calculating the transient stability limit (e.g., the critical clearing time (CCT)). This is exacerbated when considering many operating conditions, which grow significantly with the connection of intermittent RES. In addition, it is often very complicated to analyse and understand the reasons behind the changes to the stability limit using TDS – making design of targeted transient stability enhancement measures challenging. Sensitivity Analysis (SA) methods can be used to identify the impact of uncertain input variables on the variation of an output [5], which can assist in developing a more detailed understanding of the factors impacting stability (e.g., transient response prediction in [6] and small-signal stability [7]). However, SA tools are limited in that they cannot be used to reduce the computational burden of TSA.

Machine Learning (ML) has been demonstrated to be an effective way to overcome the limitations of TSA [8]–[11]. However, the barriers to implementation for ML methods to deal with TSA in real-world applications are centred around confidence in two key properties.

1) *Accuracy:* the ability of the ML model to predict the transient stability limit compared to the ground truth (typically determined via RMS TDS). This is important in establishing confidence in the ability of the ML to capture the complex power system dynamic behaviour.
2) *Interpretability:* the ability to understand how a ML model reaches predictions with respect to features (power system variables). This primarily enhances trust in the ML model (when comparing with power system engineering principles) but can also be used to increase knowledge of transient stability in a network.

A methodology adhering to both properties offers two main advantages over RMS TDS: (a) fast (and accurate) online estimation of the stability limit, and (b) increased understanding of the variables impacting the stability limit. However there exists a tension between the two, often forcing methodological trade-offs [12]. For example, linear regression and decision trees (DT) [13], [14] are straightforward to interpret but may not achieve the desired accuracy in some cases. Conversely, black-box models such as eXtreme Gradient Boosting (XGBoost) and Artificial Neural Networks (ANN) [15], [16] tend to achieve higher accuracy but are difficult to interpret. In recent years research into so-called Interpretable Machine Learning (IML) has developed [17], [18]. At the time of writing and to the extent of our knowledge, there is currently no widely accepted definition of

This work was partially supported by EPSRC Centre for Doctoral Training in Future Power Networks and Smart Grids at the University of Strathclyde and Imperial College London EP/L015471/1 (R. I. Hamilton) and by the UKRI Future Leaders Fellowship MR/S034420/1 (P. N. Papadopoulos, R. I. Hamilton).

All results can be fully reproduced using the methods and data described in this paper and references provided. For the purpose of open access, the authors have applied for a Creative Commons Attribution (CC BY) license to any Author Accepted Manuscript version arising from this submission.

The authors are with the Department of Electronic and Electrical Engineering at the University of Strathclyde, Glasgow, Scotland.

interpretability, nor is there a single accepted approach. However, authors in [19] define IML as: *"the extraction of relevant knowledge from a machine-learning model concerning relationships either contained in data or learned by the model"*. Explanation techniques can be either *local* or *global* in nature, offering insights into *single* predictions of a model or the *entire* model respectively. For TSA, local model explanations can be used to design emergency control applications and global interpretations for general planning and/or operational rules. In addition, whilst there are no formal indices relating to interpretations, techniques can be classified based on the type of insight. These are typically feature *importance* and feature *effects*. Feature effects offer insights into the impact of a feature on the model outcome, which are particularly advantageous for transient stability enhancement measure design and thus we deem them more suitable to the application proposed in this paper. Many such post-hoc IML techniques exist (reported in [20]) and authors in [21] highlight many of the pitfalls, urging caution when using IML to avoid drawing incorrect conclusions. Local Interpretable Model-agnostic Explanations (LIME) is a local technique capable of providing feature effects for individual points that can be extrapolated to form global explanations [22]. The technique was implemented for the TSA problem in [23]. A key limitation with LIME is that defining the neighbourhood around the instance for explanation is complicated and can lead to errors. LIME can also only provide additive explanations without separation of main effects and interactions [21].

Permutation Feature Importance (PFI) is a global technique [24], used in [25] to interpret DT models trained to predict the transient stability limit. PFI can provide feature importance, but not feature effects [21] and is limited in that feature importance is based on the decrease in model performance (i.e., is linked to the error of the model).

Numerous post-hoc IML methods exist, which are presented concisely in [20]. This paper proposes the use of model-agnostic Shapley Additive exPlanations (SHAP) [26], which is advantageous for three reasons. First, SHAP is a local IML technique that can be extended to form global explanations which are consistent. Second, the SHAP framework uses approximate Shapley values, which are defined as the average marginal contribution of a feature to all feature coalitions with that feature [27]. This means that feature effects are provided – in the units of the estimated quantity (a key advantage, that we later demonstrate). Shapley values have a theoretical foundation in cooperative game theory, which possess certain desirable properties that together fairly allocate the pay-out (in the case of ML, the prediction) amongst features. Thirdly, the SHAP framework proposes using a simple linear explanation model as an interpretable approximation of the complex original model. In doing so, a class of additive feature attribution methods can be defined which unifies six existing IML methods (including LIME and DeepLIFT) demonstrating that SHAP is the only method that possesses the aforementioned properties. Due to the foundations in game theory, interaction effects between features can also be analysed.

The SHAP framework has been very recently applied in the power systems domain, however not in the context of understanding the complexities of the transient stability limit, as we propose for the first time in this paper. For example, in [28] SHAP is used in the frequency stability problem. In [29], authors use SHAP values to determine the relationship between small-signal stability and topological graph metrics relating to connection of RES. SHAP is demonstrated to effectively determine the relationships between small signal stability and metrics of interest using a Random Forrest (RF) regressor. Authors in [30] integrate SHAP with dynamic security assessment for critical unit detection using classification. However, the authors seek to gain insights into a classifier and do not consider the impact of RES; which is a key driver of nonlinear changes in the dynamic response in power systems [31] that must be better understood.

This paper proposes the use of the SHAP framework to attribute an effect to power system variables on the transient stability limit (in particular, the CCT) on a locational basis. The model-agnostic nature of SHAP provides freedom regarding ML algorithm selection, enabling the prioritisation of accuracy – whilst still providing detailed insights into the stability limit. A series of location-specific ML models are trained to predict the CCT at each busbar in a network. The locational nature of the proposed method enables insights into the entire stability boundary, rather than just for a single contingency. The SHAP framework is then used to gain insights into these location-specific models, providing both local explanations of single operating points and global interpretations of the models. These global interpretations are then compared between the location-specific models to reveal regional trends in the stability boundary. In this paper, a power system variable of interest (VOI) is defined as a variable known to be important to stability, is controllable, is likely to vary and/or of interest to the user (e.g., a variable that is easily controllable). To identify the aforementioned regional trends, the covariance between a VOI and the SHAP values for that VOI from each location-specific ML model are calculated. In doing so, the impact of a change in a VOI on the entire stability limit behaviour can be assessed. The key novel contributions from this paper include:

1) the use of SHAP to understand the effects of features influencing ML models specifically trained to predict transient stability limit (specifically the CCT) at each busbar in a system along with the critical fault (the minimum CCT, $CCT_{min}$),
2) extraction of rules from SHAP values for a particular power system VOI that can lead to new detailed understanding (or reinforce existing knowledge) of complex power system dynamics,
3) identification of locational trends in the stability boundary by assessment of the covariance between a VOI and location-specific SHAP values for that variable.

## II. Methodology

Fig. 1 outlines the workflow for the proposed method. An AC OPF is used to establish generator dispatch for each operational scenario (Section II. A.). RMS TDS are subsequently executed to calculate the CCT for a fault on each busbar $n$ of the network (Section II. B.). Location-specific CCTs provide important regional information relating to the stability boundary, however the duration of the critical fault, $CCT_{min}$, is also important for protection scheme operation and so is also calculated. Section II. C. outlines the construction of the location-specific transient stability databases (TSDb) (i.e., one for each busbar $n$). Each TSDb consists of power system variables as input features, $x$, and the CCT as the target, $y$. An additional TSDb is constructed for $CCT_{min}$, resulting in $n$+1 databases. Feature selection is outlined in Section II. D. From these databases, individual ML models are trained to predict

CCT at each busbar and $CCT_{min}$, resulting in a total of $n$+1 ML models. Section II. E. outlines the ML algorithm selection process which can be guided by the desired accuracy due to the model-agnostic nature of SHAP. A particular aspect of importance – and contribution of the paper – is the focus on using ML models to reduce not just the overall prediction error, but also the minimum and maximum errors. To achieve this, black-box models are required, increasing the need for interpretability following our proposed method. This is an important aspect due to the potential system-wide impact from over- or under-estimating the transient stability limit. Accuracy metrics used in this paper are given in Section II. F.

Section II. G. summarises the SHAP framework, describing how both local explanations and global interpretations are captured. In this context, a SHAP value can be interpreted as the effect of a feature on shifting the predicted CCT from the expected baseline CCT (i.e., the CCT if the model had no training feature data) for a particular operational scenario in a particular ML model. When SHAP values across multiple operational scenarios are assessed, trends in model behaviour are captured. Since SHAP explanations are obtained for each $n$+1 location-specific model, insights of the entire stability boundary are obtained.

The method for analysing these location-specific insights and identifying locational stability trends is set out in Section II. H. Since a SHAP value is defined as the effect of a feature on the model outcome; comparing a particular feature (i.e., a VOI) to the location-specific SHAP values throughout the network reveals the system-wide effect. Locations with similar stability limit behavior to a particular VOI are identified by calculating the covariance between the actual VOI from the TSDb and the location-specific SHAP values of that variable. In doing so, the impact of a VOI on the entire stability boundary is revealed – giving an indication of locations with similar stability limit behavior to a particular variable. Since we also obtain SHAP values of a model trained to predict $CCT_{min}$, the impact of a VOI on the critical fault duration is also obtained. Finally in Section II. I. outlines PFI, which is used to demonstrate the benefits of a method capable of obtaining feature effects – as in the SHAP framework.

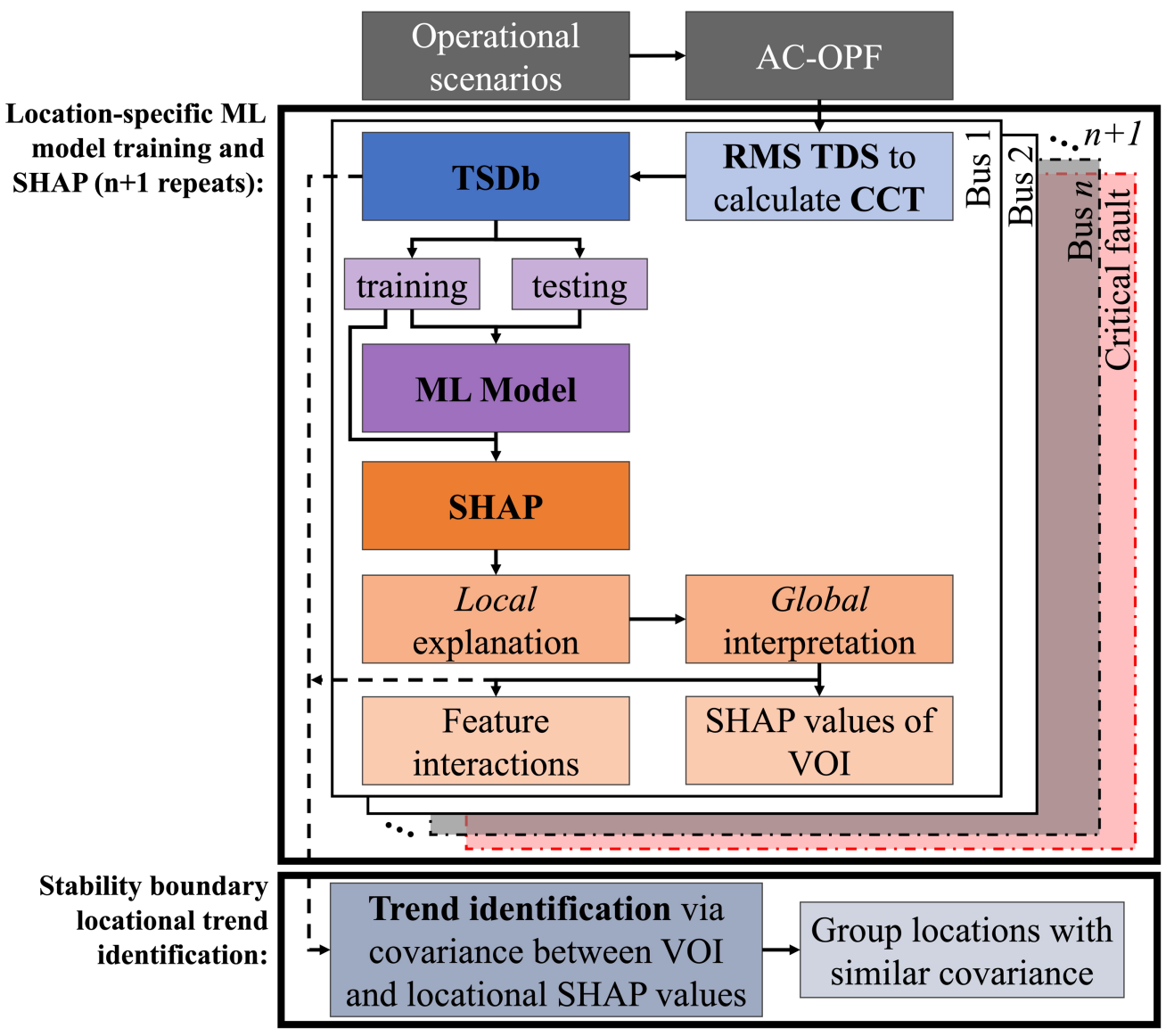

Fig. 1. Overview of proposed method for TSA using ML and SHAP.

### *A. Generator Dispatch via AC Optimal Power Flow*

An operational scenario of a network is defined as a certain realization of demand, wholesale electricity markets preference for generation dispatch, and a network state. The AC Optimal Power Flow (AC OPF) problem is used in this paper as a proxy to model the preference of a wholesale electricity system for generator dispatch, that satisfies operational constraints. The objective function for an AC OPF [32] is to find a steady state operating point that minimises the cost of generation while satisfying operating constraints and meeting demand. MATPOWER is used to perform the AC OPF and each generator's cost is modelled by a polynomial cost function [32]. The resulting dispatch is used to initialize the dynamic studies in a manner that captures how the market would dispatch generation. In doing so, the impact of the market is inherently included in the operational scenarios and thus, the stability limit.

### *B. Transient Stability and the Critical Clearing Time*

The CCT is a measure of the proximity of an operational scenario to the stability boundary for a given fault that captures the full dynamic response of the network and its generation, including converter interfaced devices. The stability limit may vary between operational scenarios and indeed between network locations [3].

The CCT for a particular fault is determined using RMS TDS and iteratively increasing the fault duration by 0.1 sec until loss-of-synchronism of a SG (or a group) is identified, then decreasing the step size to 0.01 sec to achieve the desired resolution. A limit of $\pi$ degrees on the rotor angle deviation between any two SGs for identifying loss-of-synchronism is used, mathematically described as follows;

$$\Delta\delta_{i,j} = \delta_i - \delta_j > \pi, \qquad (1)$$

where $\delta_i$ and $\delta_j$ are the rotor angles of SG $i$ and $j$. An upper limit for CCT estimation of 1.40 sec is imposed, beyond which the system is considered very far from instability.

The CCT is calculated at each busbar $n$ of the network under study for a zero-impedance self-clearing three-phase-to-ground short circuit. Busbar faults are considered in this paper since faults on a transmission line (connecting two busbars) are likely to be more transiently stable due to the increased fault impedance. In addition to $n$ locational CCTs, the $CCT_{min}$ is also determined. $CCT_{min}$ is taken as the shortest of the locational CCTs for each operational scenario and represents the minimum duration in which protection schemes must operate to maintain system stability. For this reason, $CCT_{min}$ can be used as a measure of the overall system proximity to the stability boundary for the worst-case scenario and may be of particular interest to predict and fully understand the factors that impact it. Calculation of $CCT_{min}$ is contingent on knowing location-specific CCTs and thus is computationally expensive to calculate via RMS TDS (hence this is done offline during database creation stage).

### *C. Transient Stability Database Creation*

The premise of ML methods for TSA is to train a model offline from a TSDb, capable of identifying the relationship between power system variables (features, $X$) and the transient stability of a system (target, $Y$), using a series of simulated system responses to a network fault. Each row in a TSDb represents a different operational scenario and each column a power system variable (feature). The final column in a TSDb is the target variable – the CCT (Fig. 2).

The proposed methodology seeks to understand the entire stability boundary. As such, separate ML models are trained to estimate CCT at each busbar $n$ and another for the $CCT_{min}$ – resulting in a total of $n$+1 models. Therefore, a separate

TSDb is constructed for each. Features between the series of databases are unchanged, with only the target changing (i.e., the CCT at each bus and $CCT_{min}$, $n$+1).

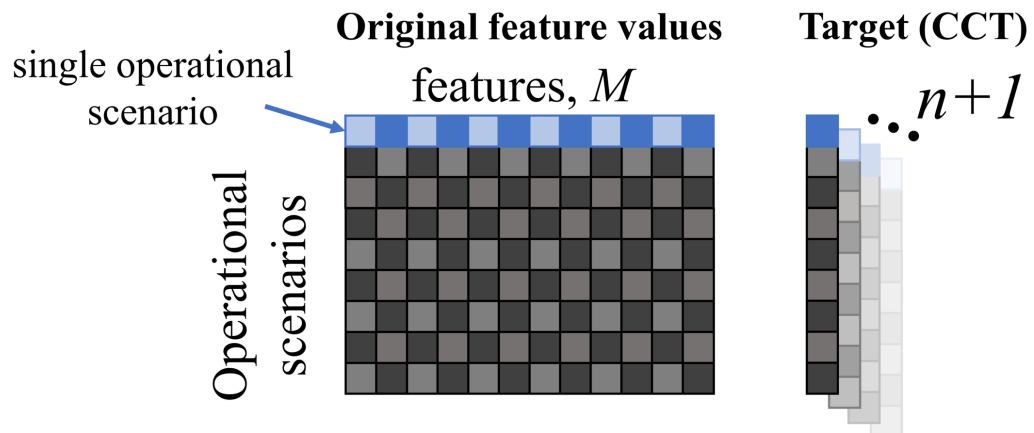

Fig. 2. Components of TSDb showing original features and different target per bus $\boldsymbol{n}$ and $CCT_{min}$, resulting in $\boldsymbol{n}$+1 databases.

*D. Feature Selection*

Feature selection in ML is important and will impact the prediction accuracy of the model (consequently dictating the types of insights provided). In addition, feature selection is important from an interpretability perspective – the primary contribution of this paper – since insights relate directly to the features selected.

Identifying the instability limit based only on pre-fault features is advantageous for preventive control actions since it allows for more time to act (e.g., generator redispatch). Therefore, post-fault data is not used as features in this paper. Instead, the features included are exclusively pre-fault and relate to SG, RES, and the power system. SG parameters include P, Q and V setpoints as well as machine size (MVA rating, inertia, and generator limits). RES parameters include P, Q and V setpoints along with the MVA rating. In addition, RES penetration parameters are included (specifically the RES MVA rating with respect to total SG MVA rating, RES MVA rating with respect to SG MVA rating per area and RES MVA rating with respect to SG active power output). System parameters include total P and Q demand, busbar voltages, busbar voltage angles and power flows on lines. The full details are provided in Table I for a network with $k$ generating areas, $v$ SG units, $w$ RES units, $l$ loads and $n$ busbars, resulting in $M$ features in the TSDb. The operational scenarios and features remain unchanged between the databases, with only the target column changing.

Feature selection can be varied by the user, based on the type of insights that are required and timescales of interest. For example, this may include the inclusion of post-fault data, however this may impact the type of application the method is suitable (since in such an instance, activation windows would be shorted).

TABLE I
FEATURE SELECTION BY CATEGORY

| SG | RES | System |
|---|---|---|
| $SG_{P,v}$ $SG_{Q,v}$ $SG_{MVA,v}$ $SG_{MVA,k,total}$ $SG_{MVA,total}$ $SG_{H,v}$ $SG_{H,k,total}$ $SG_{H,total}$ $SG_{P,max,v}$ $SG_{P,min,v}$ $SG_{Q,max,v}$ $SG_{Q,min,v}$, $SG_{Ploading,v}$ | $RES_{P,w}$ $RES_{Q,w}$ $RES_{MVA,w}$ $RES_{MVA,total}$, $RES_{MVA,total}$/ $SG_{MVA,total}$ $RES_{MVA,k,total}$/ $SG_{MVA,k,total}$ $RES_{MVA,total}$/ $SG_{P,total}$ | $P_{d,l}$ $Q_{d,l}$ $v_{bus,n}$ $v_{bus,\delta,n}$ $P_{from,n}$ $P_{to,n}$ $Q_{from,n}$ $Q_{to,n}$ |

*E. Machine Learning Algorithm Selection*

The ML algorithm is selected to achieve the desired accuracy for the estimation of CCT, following the process in Fig. 3. Accuracy is fundamental in building confidence in a model/method, since a more accurate model can represent the simulated dynamics of the power system (via RMS TDS of a representative model) and thus the actual system dynamics. However, the unnecessary use of complex models is warned against in [21]. As such, in this paper we propose increasing algorithm complexity until a predefined accuracy threshold is reached – defined in Section II. F.

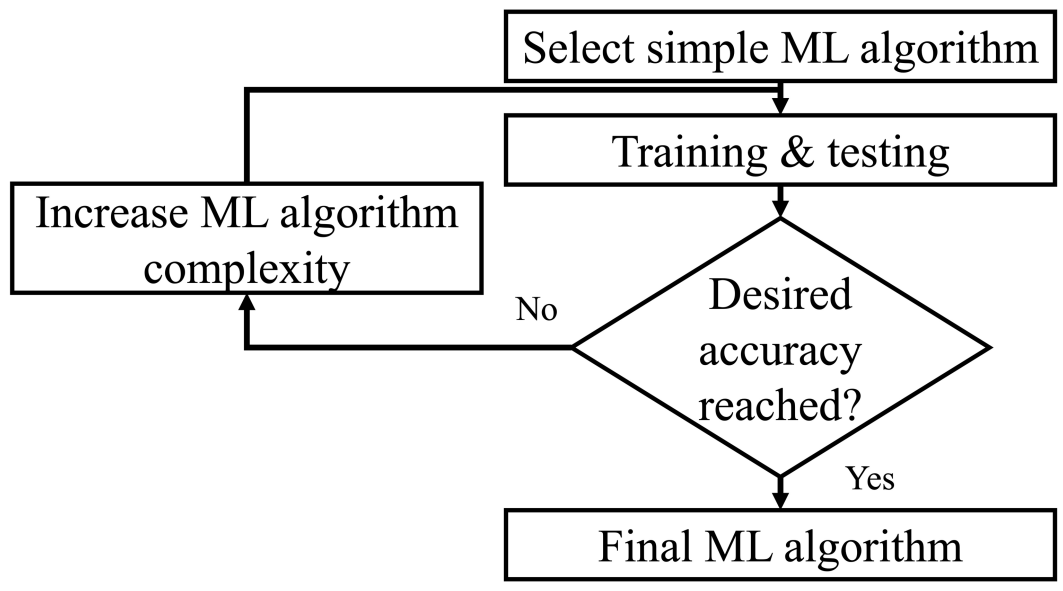

Fig. 3. ML algorithm selection process.

*F. Performance Metrics*

As previously outlined; the more accurate the ML algorithm, the less interpretable it tends to become. Moreover, an accurate model likely better represents the true system dynamics. It follows that a high degree of estimation accuracy can also offer increased confidence in the important features determined by SHAP. Therefore, in this context, accuracy is very important.

The performance of the ML algorithm will be assessed based on five performance metrics. The coefficient of determination (RSQ), mean square error (MSE) and the square root of the mean square of all errors (RMSE) (2-4).

$$RSQ = 1 - \sum_{o}^{N}(y_o - \hat{y}_o)^2 \,/\, \sum_{o}^{N}(y_o - \bar{y}_o)^2 \tag{2}$$
$$MSE = \sum_{o=1}^{N}(y_o - \hat{y}_o)^2/N \tag{3}$$
$$RMSE = \sqrt{\sum_{o=1}^{N}(y_o - \hat{y}_o)^2/N} \tag{4}$$

where $N$ is the number of data points, $o$ is the observation, $y_o$ is the actual CCT, $\hat{y}_o$ is the predicted CCT, and $\bar{y}_o$ the mean of all observations. Since the performance of different ML algorithms must be compared (Section II. E.), the average of each performance metric is reported across all $n$+1 locational models per ML algorithm.

Metrics relating to the maximum errors are also defined since an under-estimate of the CCT, may result in an exceedingly cautious generator dispatch. This is advantageous from a system security perspective but is disadvantageous from a cost perspective. Equally, an over-estimate may provide a false sense of security, potentially leading to higher instability risks. For this reason, the maximum over-estimation and under-estimation absolute errors (MOE and MUE) are determined by:

$$MOE = max\{\hat{Y}_o - Y_o\}, when\ Y_o < \hat{Y}_o, \tag{5}$$
$$MUE = max\{Y_o - \hat{Y}_o\}, when\ Y_o > \hat{Y}_o. \tag{6}$$

The MOE and MUE for the most critical faults (defined here as cases where the actual CCT<0.30 sec) are also calculated, because a large error in a more critical (shorter) fault may lead to a more serious impact. When comparing performance between ML algorithms (Section II. E.), the absolute maximum MOE and MUE of any of the $n$+1 ML models is reported per algorithm. For this application, the accuracy threshold is MOE or MUE <0.02 sec due to the importance of minimum/maximum errors in TSA. Note: this can be varied depending on user requirements.

*G. SHapley Additive exPlanations (SHAP)*

SHapley Additive exPlanations (SHAP) [26] has gained traction in ML communities, and recently in power systems [18]. This paper proposes the use of SHAP to gain insights into each locational ML model which is trained to predict the transient stability limit. In doing so, an understanding of the factors influencing the shape of the entire stability boundary can be obtained – the main focus for this paper.

SHAP is used to attribute an effect to each feature in the training database for a given operational scenario. This can be interpreted as the change in the *expected model prediction* (if no features are known, taken as the average prediction from the training database) for a feature and is in the units of the target variable. Therefore, SHAP gives the change in *expected CCT* of each feature in sec. *Local explanations* for individual operational scenarios can be analysed – from which *global interpretations* can be revealed when observing multiple scenarios. Local explanations are subunits of the global interpretations, making them consistent. This makes SHAP particularly advantageous for explanations of single operating points, but also for more general trend identification across a database of operational scenarios. In the context of TSA, local explanations are particularly advantageous in an online setting (i.e., understand the current operational scenario in detail) and global interpretations in a planning setting (i.e., understand general stability boundary trends in a network).

$$g(z') = \phi_0 + \sum_{i=1}^{M} \phi_i z'_i \quad (7)$$

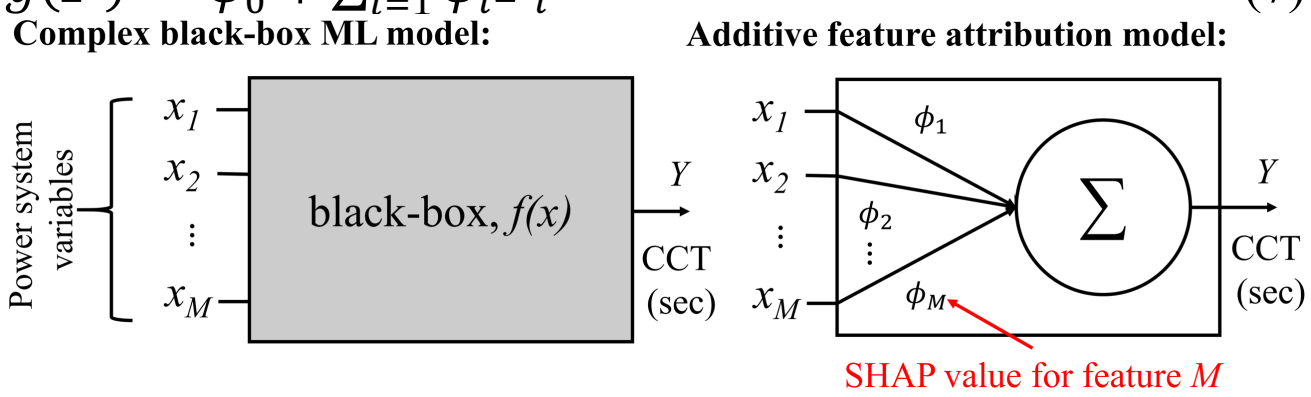

Fig. 4. Left: original model complex model (black-box). Right: simple linear explanation model (additive feature attribution model) for a particular prediction (e.g., a single operational scenario)

The SHAP framework identifies a new class of additive feature attribution methods [26]. Such methods seek to build a simpler *explanation model* $g$ to explain the complex *original model* $f$ (which is often black-box), as shown in Fig. 4. This explanation model is expressed as a linear function of binary variables that enables explanations of a model prediction based on a single input $x$. Simplified inputs $x'$ are used to map to the original input via some mapping function $h_x$, where $x = h_x(x')$. The goal is to ensure $g(z') \approx g(h_x(z'))$, whenever $z' \approx x'$ and where $z' \in \{0,1\}^M$ and $M$ is the number of simplified input features. The method attributes an effect $\phi_i$ to each feature (where $\phi_i \in \mathbb{R}$) and summing the effects of all feature attributions approximates the output $f(x)$ of the original complex model (7). This results in an intuitive feature importance measure. The framework demonstrates that for additive feature attribution methods to adhere to three desirable properties Shapley values from cooperative game theory [27] must be used (and thus improves previous methods) [26]:

1) *local accuracy*: when approximating the original model $f$ for a specific input $x$, the explanation model $g(x')$ should at least match the output of $f(x)$ for the simplified input $x'$ (which corresponds to the original input $x$ via $x = h_x(x')$),
2) *missingness*: if the simplified inputs represent feature presence, then features missing in the original input should have no impact $x'_i = 0 \Rightarrow \phi_i = 0$,
3) *consistency*: if a model changes so that some simplified input's contribution increases or stays the same regardless of the other inputs, that input's attribution should not decrease. Let $f_x(z') = f(h_x(z'))$ and $z' \backslash i$ represent fixing $z'_i = 0$. For two models $f$ and $f'$, if $f_x{}'(z') - f_x{}'(z' \backslash i) \geq f_x(z') - f_x(z' \backslash i)$ for all inputs $z' \in \{0,1\}^M$, then $\phi_i(f', x) \geq \phi_i(f, x)$.

The unified approach proposed by SHAP improves previous methods by preventing unintentional violations of these properties. A Shapley value can be interpreted as the average marginal contribution by a feature to all coalitions that contain contributions from this feature, meaning that SHAP is more than just a sensitivity-based approach. The Shapley value, $\phi_i$, of feature $i$ is given by:

$$\phi_i(f, x) = \sum_{z' \subseteq x'} \frac{|z'|!(M - |z'| - 1)!}{M!} [f_x(z') - f_x(z' \backslash i)], \quad (8)$$

where $f_x(z') = f(h_x(z'))$ and $z' \backslash i$ denotes setting $z'_i = 0$. Note: if we let $\phi_0 = f_\emptyset(\emptyset)$ then the Shapley values match (7) and is hence an additive attribution method. SHAP requires a model comparison which is retrained on the features that are not withheld – which is hard in a ML context. This is overcome by expressing the explanation model as a "conditional expectation function of the original model" [26] (i.e., the *explanation model* in the SHAP framework is a conditional expectation function of the *original model*). Therefore, SHAP values are the solution to (8), where $f_x(z') = f(h_x(z')) = E[f(z)|z_S]$ and $S$ is the set of non-zero indexes in $z'$. When conditioning on a feature, SHAP values attribute the change in model prediction to that feature, explaining how to move from the base value $E[f(z)]$ (i.e., the prediction if no features are known (i.e., $\phi_0 = f_\emptyset(\emptyset)$), which is taken to be the average of all predictions in the training database) to the current output $f(x)$. Therefore, in the context of this paper a SHAP value gives the average *effect* – in *sec* – of a VOI taking a particular value on shifting the CCT from the expected value (average CCT in the training database) to the predicted CCT.

Shapley values are computationally expensive to compute because the number of coalitions increases exponentially with the number of features, and thus becomes intractable. SHAP overcomes this by introducing various so-called explainers, some of which calculate approximate Shapley values by leveraging ML algorithm intricacies to reduce computational time (using fewer coalitions of features – see details of various methods in [26]). For example, KernelExplainer is model agnostic but computes approximate values in exponential time. TreeExplainer can only be used with tree-based algorithms but calculates exact values in polynomial time. DeepExplainer (used in this paper) is specific to neural networks and calculates approximate SHAP values based on user-defined background samples, with computational time scaling linearly with the number of samples. Therefore, there exists a trade-off between ML algorithm accuracy and speed, and SHAP implementation accuracy and speed that must be considered in design stages.

The SHAP framework provides explanations of the ML model. Therefore, if the underlying ML model is inaccurate, the SHAP explanations will reflect this. This highlights the need for the use of highly accurate ML models. Moreover, it is important to note that while SHAP values can reveal associations learned from the data, they do not necessarily guarantee or reflect causal relations. Identified associations can be helpful for system operators and planners but would need further validation based on domain knowledge or other causal inference methods for causal effects to be guaranteed.

Local explanations give a 1D matrix of SHAP values of each feature, $i$. Global interpretations can be obtained by repeating this across the entire training database; resulting in a 2D matrix of SHAP values consisting of one row per operational scenario and one column per feature, $i$ (Fig. 5). Since in this paper we generate $n$+1 ML models, a 3rd dimension is added (Fig. 5). The resultant 3D matrix therefore consists of the SHAP values for all features across all operational scenarios in the training database for each $n$+1 locational ML models. This matrix is later used in Section II. H. for stability boundary trend identification.

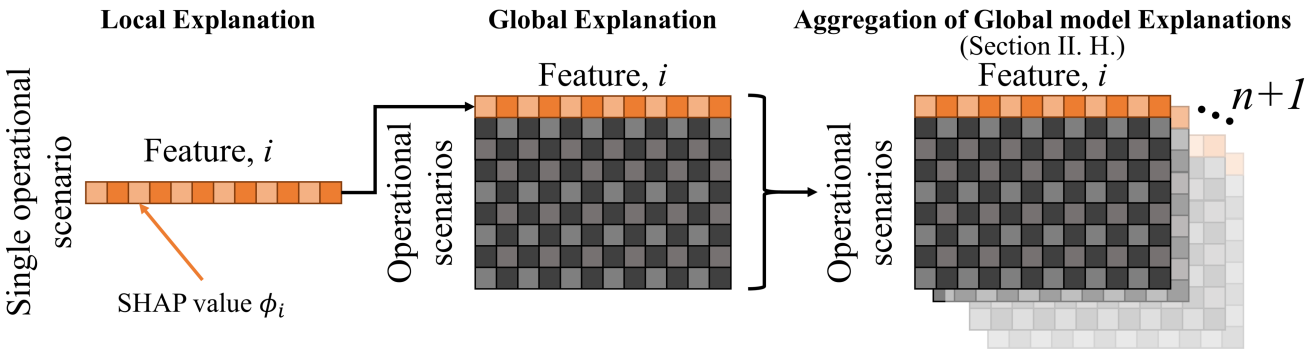

Fig. 5. Data structure for SHAP local, global and for locational trend identification.

## *H. Identification in Locational Trends in SHAP Values via Covariance*

This paper seeks to identify how a change in a particular power system VOI may impact the stability limit throughout the network. To do so, covariance (a measure of how two variables will change together) between a VOI and the location-specific SHAP values corresponding to that VOI for each ML model is calculated – resulting in $n$+1 covariance values per VOI. In doing so, the effect of a VOI across the entire network is obtained – capturing locations with 'similar' (or 'dissimilar') stability limit behaviour with respect to that VOI.

Fig. 6 shows how the covariance between the *actual* feature values from the TSDb for a given VOI and the corresponding column for that VOI in the 3D matrix of SHAP values for each model is calculated. This results in a sense of how the stability boundary will be impacted due to changes in some VOI. Such insights may inform design and development of stability enhancement measures.

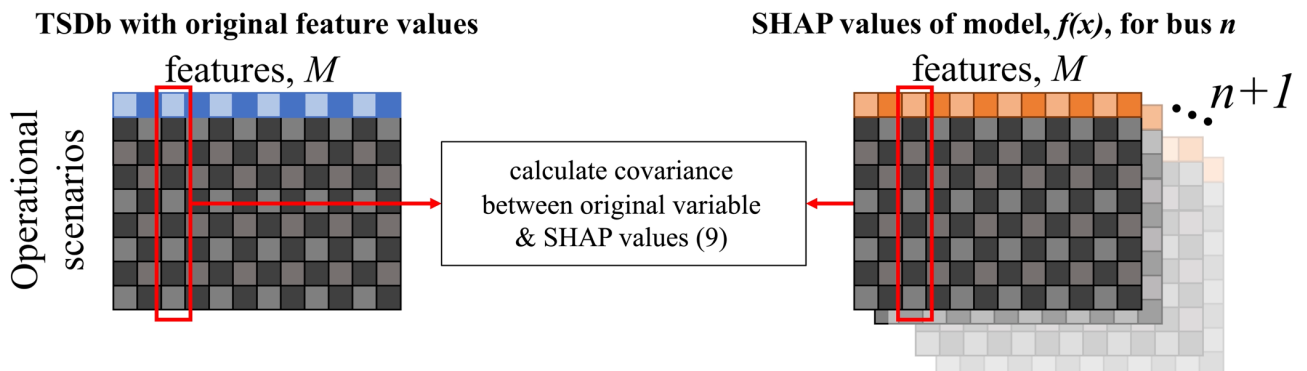

Fig. 6. Identification of locational trends between actual VOI value and the corresponding SHAP values from each ML model.

Covariance describes how two variables are related to one another. That is, a measure of how two random variables in a data set change together. This is described mathematically as:

$$COV(a,b) = \sum_{i=1}^{N}(a_i - \bar{a})(b_i - \bar{b})/N - 1 \qquad (9)$$

where $a$ is the independent variable (in this case the power system VOI), $\bar{a}$ the mean of $a$, $b$ the dependent variable (SHAP value for the VOI at a given location), $\bar{b}$ the mean of $b$, and $N$ represents the number of data points in the sample. This is repeated for each location, resulting in $n$+1 covariance values per VOI. Positive covariance means that the two variables are positively related, moving in the same direction. A negative covariance means that the variables are inversely related, moving in opposite directions. Observing covariance at between the VOI and SHAP values at each $n$+1 location reveals the impact of a VOI on location-specific SHAP values and thus on the transient stability limit throughout the network.

## *I. Permutation Feature Importance*

In [25], authors propose the use of PFI [24] as the IML technique to extract understanding of the transient stability limit. PFI is limited in that the output is a ranked list of features based on the mean importance and the decrease in model performance when a feature is permutated. In this paper, we compare SHAP with PFI to illustrate the advantage of SHAP in providing feature effects. Whilst useful for identifying important features; interpreting the meaning of the importance score has no relation to the effect on the target variable. For model $f(x)$ of the dataset $X$ with target $Y$, the reference score is computed based on error measure $L(Y, f(x))$. For $k$ repeats in $k = 1, \dots, K$, each feature $i = 1, \dots, M$ in $X$ is randomly shuffled to generate a corrupted version of the data $\bar{X}_{k,i}$. The score of the model with the corrupted data ($score_{k,i}$) is computed, and importance $imp_i$ of feature $i$ given by:

$$imp_i = score - {}^{1}/_{k} \sum_{k=1}^{K} score_{k,i}. \qquad (10)$$

## *J. Implementation Details in a Realistic Setting*

Two potential applications can be envisaged for the proposed method: an online setting (in operational timescales) and offline setting (in planning timescales). In an online setting, measurement data is required that can be gathered from Phasor Measurement Units (PMUs) and Supervisory Control and Data Acquisition (SCADA) systems. In this setting, ML models are trained offline using a database of previous operational scenarios encountered. In operational time (e.g., per settlement period), the current operational data are used as inputs to the ML models to get stability predictions. Local SHAP explanations reveal the factors impacting the individual operational scenario, enhancing the operators understanding of the factors impacting the stability limit and to what extent. Cases with short CCTs are likely to be of particular interest, where SHAP explanations can be useful in guiding actions to improve system stability.

In an offline setting, the use of live measurement data is not required. Global SHAP interpretations can be used to analyse general stability boundary trends. Such insights may be used to inform general operational planning decisions (e.g., generator export or line flow limits) and/or planning decisions (e.g., construction of new assets). This can improve operational knowledge enabling the operator to avoid or mitigate conditions that may lead to poor stability margin.

## III. Test Network and Case Study Details

The proposed method is demonstrated on an adapted version of the IEEE 39-bus test network (Fig. 7). The RMS TDS simulations are carried out in DIgSILENT PowerFactory [33] and the AC OPF in MATPOWER [34]. A set of operational scenarios, designed to represent how the generation connected to the system may change because of planning or operational decisions (i.e., closure of SG plant and the connection of RES), is defined in Section III. B.

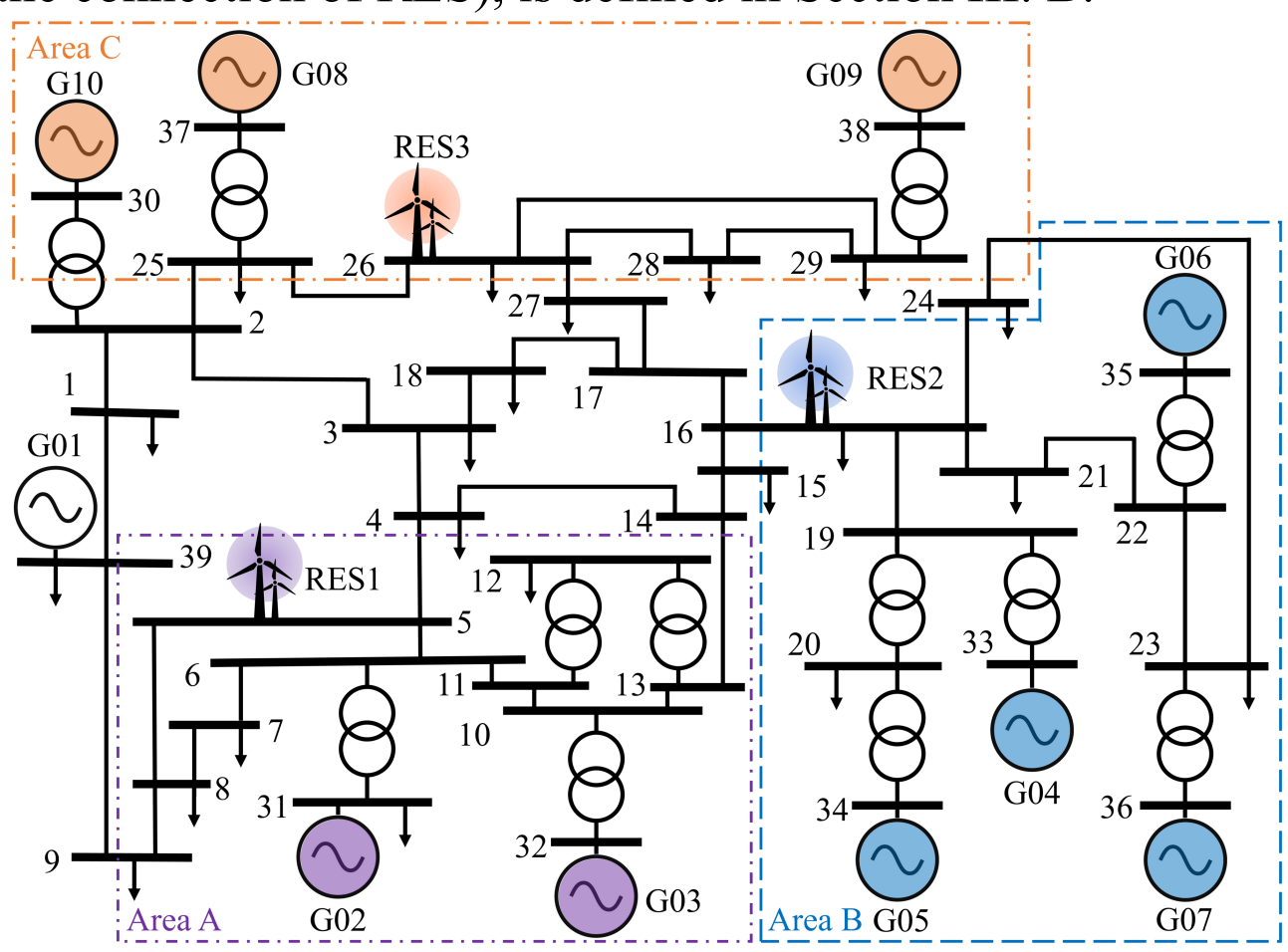

Fig. 7. Adapted version of the IEEE 39-bus network showing defined generation areas and locations for wind connection.

### A. Network Overview and Operational Scenarios

The network nominal voltage is 345 kV at a nominal frequency of 60 Hz. The total active power demand of the network is 6097.1 MW, modelled as balanced three-phase constant impedance loads in the dynamic simulations. SGs (G01-G10) are modelled as four-winding 6th order models [34]. The slack generator is taken to be G01 without an AVR or PSS and represents the '*rest-of-system*' and is not disconnected in any scenario. These machines are taken to represent four equal-sized units to simplify disconnection stages in case studies. International Electrotechnical Commission (IEC) Type-4A wind turbines [35] are used to model wind generation in these studies. A windfarm is treated as an aggregate of individual 2 MW turbines connected in parallel, each with its own transformer. The volume of wind and its location on the network depends on the operational scenario. Three generation areas are defined (Fig. 7) and the network configuration remains unchanged throughout.

### B. Generation of Operational Scenarios and AC OPF

The range of operational scenarios considered are designed to represent operating conditions that may arise due to displacement of conventional SG (as fossil-fuel used less frequently or retired) by RES (as windfarms are constructed). Three factors are varied to generate the operational scenarios: SG connected, RES connected and system demand. Specifically, one SG is displaced by RES connecting in the same generating area at a time.

$$SG_{MVA,new} = u\,(SG_{MVA,old}/4) \tag{11}$$

$$RES_{MVA} = r\left(\left((5-u)\,SG_{MVA,old}/4\right) + s\left(SG_{MVA,old}\right)\right) \tag{12}$$

Since each SG is assumed to represent four equal-sized units, each SG is disconnected in four stages (11). The new MVA rating of $SG_{MVA,new}$ is based on the number of remaining generating units, $u$ (where $u$ = 1, 2, 3, 4 and is rated to the original machine rating, $SG_{MVA,old}$).

The capacity of RES connecting to the network is scaled in a similar, inverse, manner (12). However, an additional scaling factor, $s$ (where $s$ = 0, -0.05, 0.05 in this paper), is included to decouple any proportional relationship between decreasing SG MVA rating (and subsequent reduction in inertia) and increasing RES MVA rating. In addition, a scaling factor, $r$ (where $r$ = 1, 1.4 in this paper, representing high and very-high renewable penetration respectively), is applied to account for very high RES output scenarios. RES is rounded up to the nearest even value since each RES unit is rated at 2 MVA. Scenarios with no displacement are also considered. Total system demand is varied uniformly throughout the network from 0.6 to 1.025 p.u., based on the initial power flow of the network (given in [33]), in 0.025 p.u. increments.

3906 operational scenarios are designed based on 18 demand levels, 36 SG disconnection stages (9 SGs, each disconnected in 4 stages) and 6 RES levels. This results in 3888 displacement scenarios plus an additional 18 scenarios with no displacement.

A full AC OPF is implemented in MATPOWER to establish generator dispatch for each operational scenario (Section II. A.). Cost coefficients representative of different generation types are taken from [36] and allocated to generators as in [37], with RES is considered to have no marginal cost. Of the 3906 operational scenarios; 3762 (96.3%) successfully converge, with the remaining 144 not converging. The successfully converged scenarios are passed to DIgSILENT PowerFactory, where RMS TDS are executed to determine the CCT for all fault locations. As specified in Section II. B., fault locations considered in this case study are three-phase-to-ground busbar faults excluding both the LV terminal busbars for all SGs (9 busbars in the 39-bus network Fig. 7) and all RES points of common coupling (3 busbars), and bus 1 and 9 (2 busbars, due to their highly stable nature). This results in 25 fault locations (i.e., $n$=25). Each location-specific TSDb therefore consists of a total of 223 power system variables (i.e., $M$=223, described in Section II. D.) which is subsequently used in model training.

## IV. Results

The following section details key results and findings from implementing the proposed methodology on the IEEE 39-bus test network for the case study previously described – with a primary focus on interpretability. In Section IV. A, results and discussion related to the accuracy of the 26 ($n+1$) ML models are presented. Section IV. B provides details on the computational speed of the method. In Section IV. C showcases results for a local SHAP explanation (i.e., for a single operational scenario). In Section IV. D a global PFI interpretation is provided as a benchmark for interpretability. Results are then compared with a global SHAP interpretation in Section IV. E to demonstrate the benefit of this unified approach with foundations in cooperative game theory. Finally, system-wide stability boundary trends are identified in Section IV. F.

### A. Accuracy

The accuracy of three different tree-based algorithms are compared – specifically DT, RF, XGBoost and a feedforward ANN with 2 hidden layers (Section II. E). Accuracy of each algorithm is reported in Table II using the metrics outlined in Section II. F. Results here show that performance improves as the algorithm becomes increasingly opaque. The high degree of accuracy achieved by the black-box ANN for locational CCT estimation increases confidence in the interpretation of the model, since the locational models are closely replicating RMS TDS results – compared to DT, RF and XGBoost algorithms. In particular, the ability of ANN to reduce outliers is highly advantageous over the other algorithms. This is important for transient stability applications, where the consequences of a large over-estimate of the stability limit could result in a loss-of-synchronism, a cascade and ultimately a black out. Therefore, based on the method outlined in Section II. E, the ANN algorithm is used and any further increases in model complexity deemed unnecessary.

TABLE II
Accuracy Metrics for all ML Models

| Performance Metric | DT | RF | XGBoost | ANN |
|---|---|---|---|---|
| Average RSQ | 0.9731 | 0.9815 | 0.9936 | 0.9932 |
| Average MSE (sec$^2$) | 0.0013 | 0.0008 | 0.0003 | 0.0002 |
| Average RMSE (sec) | 0.0288 | 0.0228 | 0.0137 | 0.0110 |
| Max MOE (sec) | 0.37 | 0.29 | 0.14 | 0.08 |
| Min MUE (sec) | 0.32 | 0.26 | 0.13 | 0.07 |
| Max MOE < 0.3 (sec) | 0.14 | 0.11 | 0.03 | 0.01 |
| Min MUE < 0.3 (sec) | 0.01 | 0.01 | 0.01 | 0.01 |

### B. Computational Speed

The methodology was conducted on an Intel CoreTM i7-6700 CPU @3.40GHz with 16 GB installed RAM. An average of 40 sec are required to calculate the CCT for a given contingency using a Python script to automate DIgSILENT PowerFactory. A single ANN can be trained in approximately 60 sec in Python. All above stages are conducted offline when time is less expensive. CCT estimation in an online setting is however time critical, with inference of a single point taking approximately 0.2 sec (200-fold time reduction compared to RMS TDS). SHAP explanations from the training portion of the TSDb are constructed offline (using the DeepExplainer package in Python [26], taking under 1 sec using this PC specification). Note that the ML algorithm (which may vary with accuracy requirements as in Section II. E) and SHAP explainer selected will impact computational speed.

The method is tested on a well-known test system to simplify analysis and focus on the key interpretability theme. However, accurate estimation of transient stability status using larger systems has been proven in [38], [39]. For the proposed method to be implemented in larger systems, the additional computational burden primarily relates to the increased number of RMS-TDS required to construct ML models at each busbar and thus capture the entire stability boundary of a system. Although this is conducted offline, should the computational burden of this need to be reduced, this can be achieved by (a) reduction in number of busbars required (and thus number of RMS-TDS and ML models required), (b) increase computational power (using multiple cores, or multiple PCs to parallelize tasks).

With respect to SHAP implementation and scaling up in larger networks, for DeepExplainer implemented in this paper time scales linearly with the number of observations used. Consequently, reducing the number of background samples (for example when scaling up in large networks) would reduce computational time, but also potentially reduce accuracy of SHAP values. Note that in this paper we use the entire training set as background data.

### *C. Local SHAP Explanation: $CCT_{min}$ Model*

In this section, a local explanation of the power system variables that impact the estimation of the critical fault ($CCT_{min}$) for a particular operational scenario is provided. Similar analysis of the remaining 25 ML models (that predict the CCT at each busbar) can also be conducted but is omitted from this paper due to space constraints.

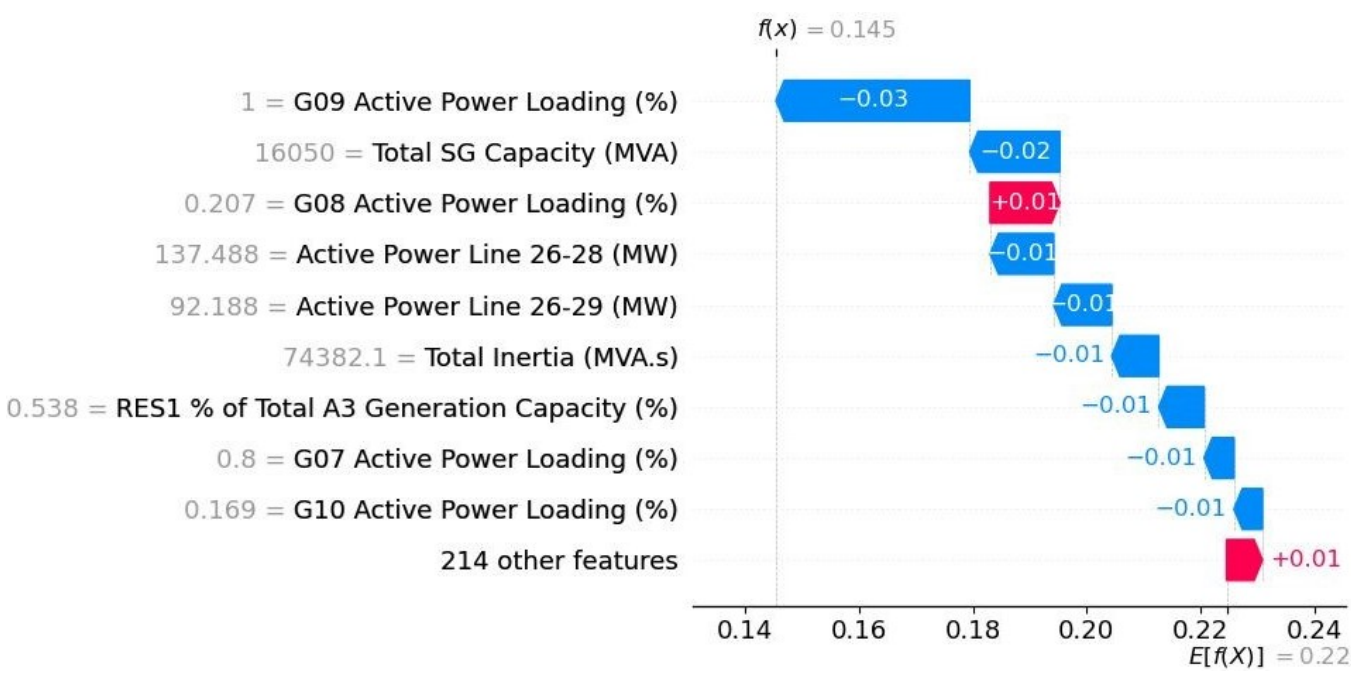

Fig. 8. Example of a local SHAP explanation of the $CCT_{min}$ model for a single operational scenario showing feature SHAP values, $\boldsymbol{\phi_i}$.

One such operational scenario example is provided in Fig**.** 8, where the base value ($E[f(x)]$) is 0.225 sec (x-axis). All features included in the TSDb are on the y-axis, ranked from top to bottom based on feature effect (the magnitude of the SHAP value) for this specific prediction. The actual value of the feature is also given on the y-axis. The impact of each feature taking that value on the prediction is represented by a blue or pink arrow and corresponding SHAP value. This value is the attributed effect on the model prediction of that feature. Moving upwards from the base value ($E[f(x)]$), the impact of each feature shifts the prediction, until the actual model estimate $f(x)$ is reached. Here the final prediction of $CCT_{min}$ for this operational scenario is 0.145 sec. The sum of the SHAP values is equal to the difference between the actual prediction and the expected value, demonstrating the additive nature of SHAP, first introduced in (7).

Starting at the base value ($E[f(x)]$) of 0.225 sec, the lowest ranked 214 features only have a combined impact of +0.01 sec on the $CCT_{min}$ prediction. G08 active power loading (3rd most important feature) in this operational scenario is 20.7%, which consequently increases the model prediction by 0.01 sec. Conversely, the impact of G09 active power loading (the highest-ranked feature) being 100% has a -0.03 sec impact on the $CCT_{min}$ prediction. These types of insights can be compared to engineering knowledge to enhance confidence in the results from the ML model and indeed SHAP. In an online setting, for example in a control room, this type of information could be used to support control room engineers – either through highlighting additional power system variables to consider or reinforcing engineering judgements. Similar local explanations for all operational scenarios can be generated, providing a global interpretation of the model, as demonstrated in the proceeding section.

Local explanations are of particular use in an online setting, for cases with relatively short CCTs. The details provided can be used to inform potential actions that will improve the CCT for that scenario. Based on the SHAP explanation in Fig**.** 8, a system operator has several potential options available to improve $CCT_{min}$. One such option would be to decrease the G09 active power loading.

### *D. Global Interpretation via PFI: $CCT_{min}$ Model*

A global explanation of the power system variables that impact the estimation of the critical fault ($CCT_{min}$) is provided using PFI (following the method presented in [25]). A similar analysis is conducted for the remaining 25 ML models. As seen in Fig. 9, PFI provides feature importance, but no feature effects (a key disadvantage, which SHAP overcomes).

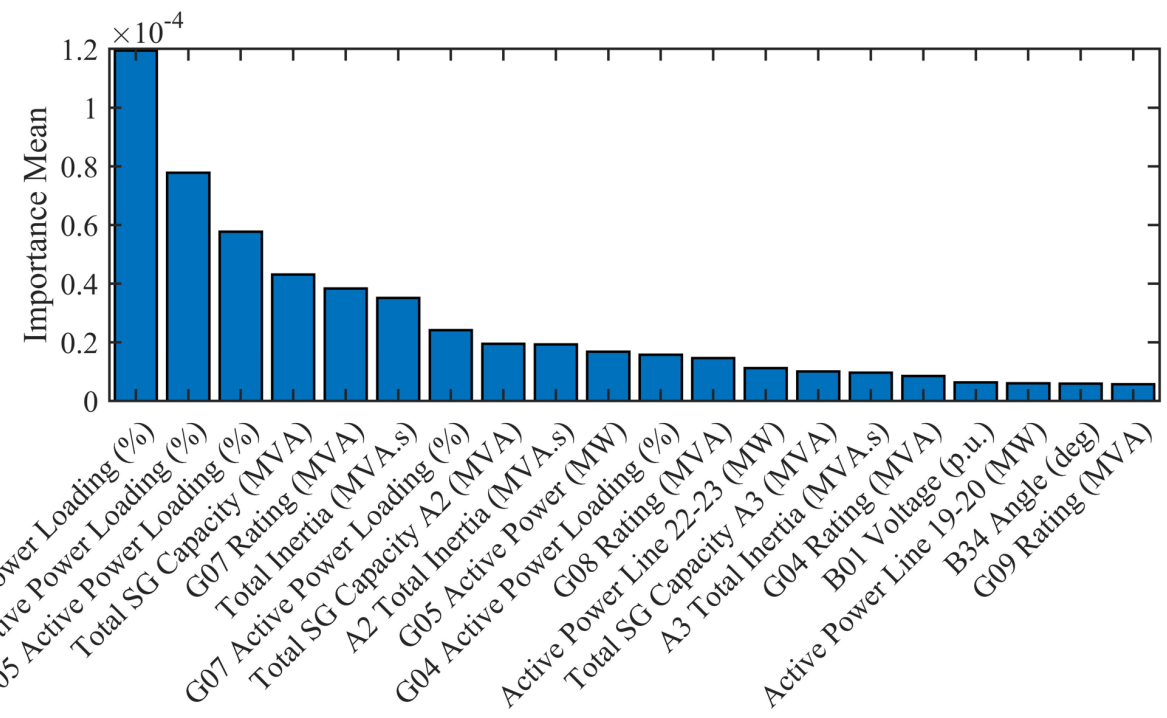

Fig. 9. Permutation feature importance (PFI) showing importance-ranked list of the top-20 features for $CCT_{min}$ estimation.

### *E. Global Interpretation via SHAP: $CCT_{min}$ Model*

As previously outlined, a global interpretation of a model can be achieved by extending local SHAP (such as the one presented above in Section IV. C.) across all operational scenarios. This results in a matrix of SHAP values (Fig. 5), with one row per operational scenario and one column per feature. This can be presented in a SHAP summary plot (e.g., Fig. 10), that provides details relating to the impact of important features (and their magnitude) on the model estimate for all operational scenarios.

The SHAP summary plot for $CCT_{min}$ (Fig. 10) provides an indication of the relationship between top-20 important features and their values on the estimation of $CCT_{min}$. Each point on Fig. 10 is a SHAP value of a feature for a given operational scenario. The position on the y-axis is determined by the feature importance (mean SHAP value across all scenarios) and on the x-axis the SHAP value (i.e., the impact of that feature on the model prediction for that particular operational scenario). The color represents the feature value for that operational scenario from *'low'* (blue) to *'high'* (pink) based on the original feature value from the training portion of the TSDb. Any point with a positive SHAP value indicates an improvement in the $CCT_{min}$ prediction from the base value (E[f(x)], which is equal to average $CCT_{min}$ for all training samples). Conversely, any datapoint with a negative SHAP value indicates a decrease in the estimation of $CCT_{min}$ from the base value. Overlapping points are distributed in y-axis direction, providing a sense of the distribution of the SHAP values per feature.

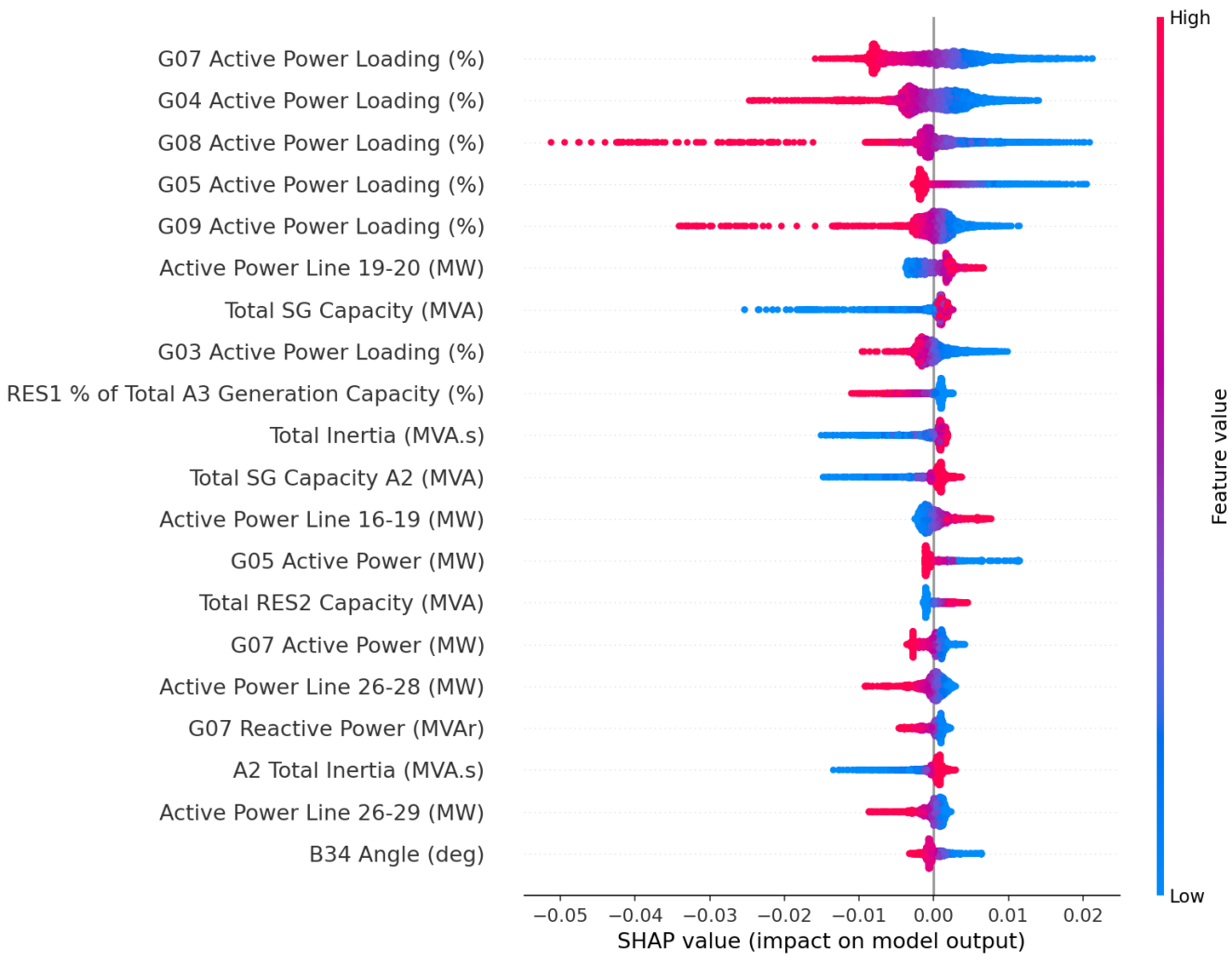

Fig. 10. Global SHAP summary plot for $CCT_{min}$ model; showing feature importance, SHAP values and feature values for the top-20 most important features for $CCT_{min}$ estimation.

There is some degree of consistency between PFI (Fig. 9) and SHAP (Fig. 10) feature importance, with 12 out of the 20 top important features appearing in both importance lists. The differences between the ordering of feature importance relates to the key differences between the methods; with SHAP accounting for interaction effects between features whilst PFI only accounts for the impact on model score with and without that feature (i.e., no interactions captured). In addition, the level of detail able to be extracted is significantly greater with SHAP, since it provides the feature *effects* and PFI does not.

The top-5 features in the example provided in Fig. 10 relate exclusively to the active power loading of specific SGs, which is extensively reported to be important for transient stability in the literature [40]. Whilst this can be considered trivial from a power system dynamics perspective, the fact that important features match up with domain expertise enhances confidence in the ability of the black-box model to accurately represent power system dynamics. In addition, SHAP is useful in that specific generators can be pinpointed and the extent of the impact on the CCT prediction determined for a large range of operational scenarios – which cannot be easily achieved using RMS TDS alone. For example, Fig. 10 pinpoints G07 active power loading (%) as the most important variable for $CCT_{min}$ estimation. When G07 active power loading (%) is *'low'* (blue), the model estimation for $CCT_{min}$ increases from the expectation by up to 0.02 sec. Conversely when G07 active power loading (%) is *'high'* (pink), the $CCT_{min}$ model estimation decreases by up to 0.015 sec. Since a common color-axis is used for all features, the summary plot in Fig. 10 is unable to provide precise details of the actual generator active power loading due to the common color axis used (however this can be plotted independently as demonstrated below).

The impact of other VOI can also be observed, providing detailed information on the complex dynamics affecting transient stability for example; '*low'* SG capacity results in a *reduction* in $CCT_{min}$ estimation. Similarly, *'high'* RES2 capacity results in an *increase* in $CCT_{min}$ estimation. The resulting insights depend on the features included in the TSDb, highlighting the importance of careful and considered feature selection. These detailed insights can be used to develop planning and operational planning rules, as we demonstrate in the following sections. These rules can help system planners and operators avoid conditions that lead to a poor transient stability limit.

*1) Detailed feature dependence insights:* While the SHAP summary plot highlights high-level relationships between power system variables and their impact on the prediction (as described above), additional insights can be provided by looking at specific VOI in detail – which can be achieved using a dependence plot (e.g., Fig. 11) that reveals the effect a single feature has on the predictions made by the model.

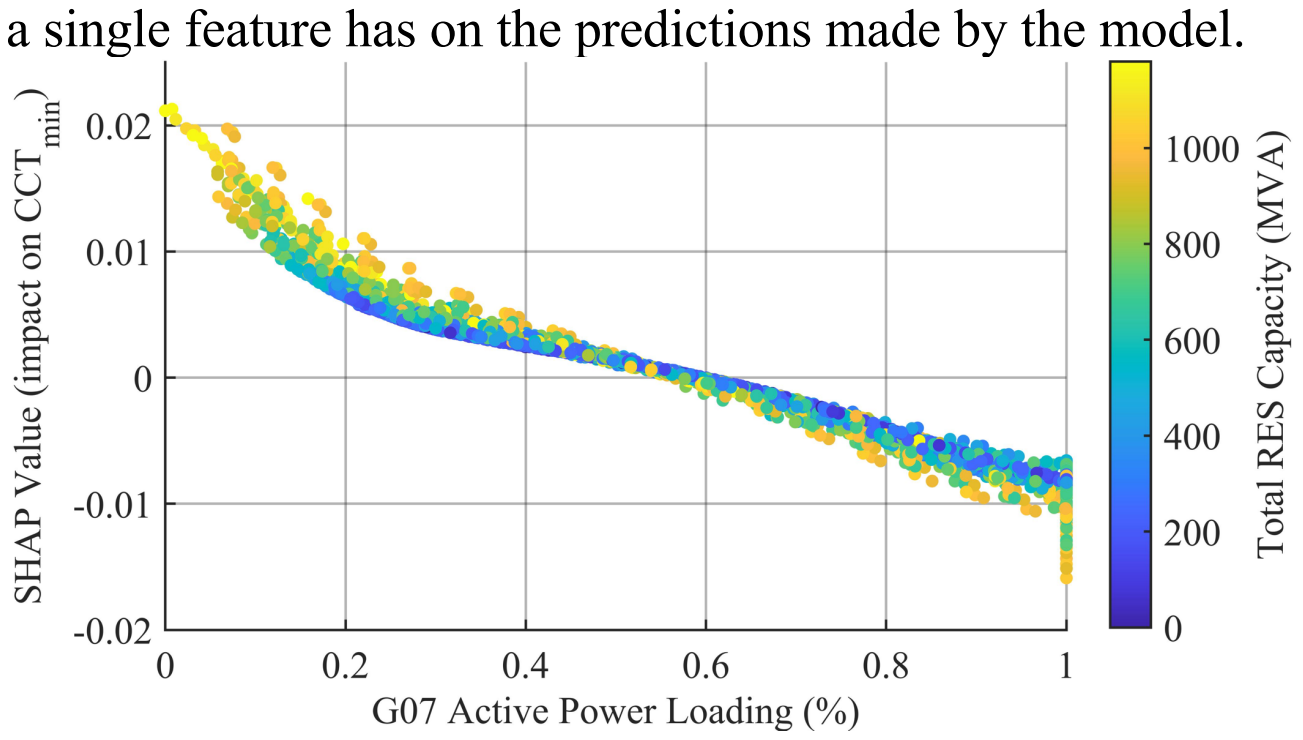

Fig. 11. Dependence plot for G07 active power loading (%) vs. SHAP value (the impact on $CCT_{min}$ prediction), with total RES capacity color axis.

In a dependance plot, the SHAP values for a primary VOI can be plotted on the y-axis, with the feature value on the x-axis. An additional secondary VOI can be added to the color axis, capturing interaction effects between features. In Fig. 11, G07 active power loading (%) is taken as the primary VOI, since it is found to be the most important variable for $CCT_{min}$ estimation (Fig. 10). The corresponding SHAP values are given on the y-axis. The secondary VOI – here taken to be the total RES capacity (MVA) installed (since it may be of interest to assess the overall impact of RES on the transient stability limit) – is plotted on the color axis. When G07 active power loading <55%, the impact on the model estimation of $CCT_{min}$ is positive (i.e., increases the prediction for $CCT_{min}$). Inversely, when G07 active power loading >55%, the impact on the model estimation of $CCT_{min}$ becomes negative (i.e., reduces the estimation for $CCT_{min}$). Therefore, an operational rule may be: *"maintain G07 active power loading (%) below 55%".* Since the SHAP values account for coalitions of features (Section II. C), such rules account for the impact of the other features. The vertical distribution of points in Fig. 11 for a given value on the x-axis highlights that there are other variables impacting the $CCT_{min}$ prediction. For example, when G07 active power loading is 100% the SHAP values range between -0.006 and -0.016.

In addition to this, the impact of the total RES capacity (MVA) on $CCT_{min}$ estimation with respect to G07 active power loading (%) is revealed. The largest improvement in the $CCT_{min}$ prediction is when G07 active power loading (%) <55% and the system has a large RES capacity (MVA) connected. However, this trend reverses when G07 active power loading (%) >55% – suggesting an interaction effect between the two variables. Therefore, another layer of complexity can be added to the previously described operational rule: *"maintain G07 active power loading (%) below 55%, particularly if there is a high capacity of RES connected to the system"*. Identification of such relationships between power system variables on the stability limit is a key benefit of using SHAP and can be extended to other VOI.

### *F. Locational Trend Identification using SHAP Values*

A key aspect of the proposed methodology is the ability to identify trends not only between power system variables and the CCT at a certain location, but also between locations – outlined in Section II. H. Two examples are given below.

*1) Example A; system-wide impact of SG active power loading:* Taking the critical fault ($CCT_{min}$) once again as an example, a reasonable approach to enhance the stability limit would be to decrease G07 active power loading (most important feature in Fig. 10). The impact on CCTs of taking such an action can be assessed by calculating the covariance between G07 active power loading and the SHAP values for G07 active power output at each location, i.e., for the different ML models trained for each location (Fig. 12). Locations B15, B17, B18 and B21-B24 all have similar (negative) covariance to $CCT_{min}$, suggesting that a *decrease* in G07 active power output will result in stability limit *improvement* at these locations (i.e., inversely related). B14 has a small positive covariance, indicating CCT at this busbar will *deteriorate* (i.e., shorten). Fig. 13 illustrates this for $CCT_{min}$, B15 and B14. A reduction in the active power loading of G07 from 100 to 40% may have as much as a 0.04 sec increase in $CCT_{min}$, 0.25 sec improvement at B15 – whilst reducing B14 CCT by 0.02 sec. Since the typical CCT range for $CCT_{min}$ is much shorter (more critical) than the typical CCT range for a B14 fault (Fig. 14), this local deterioration in stability limit may be deemed acceptable on balance.

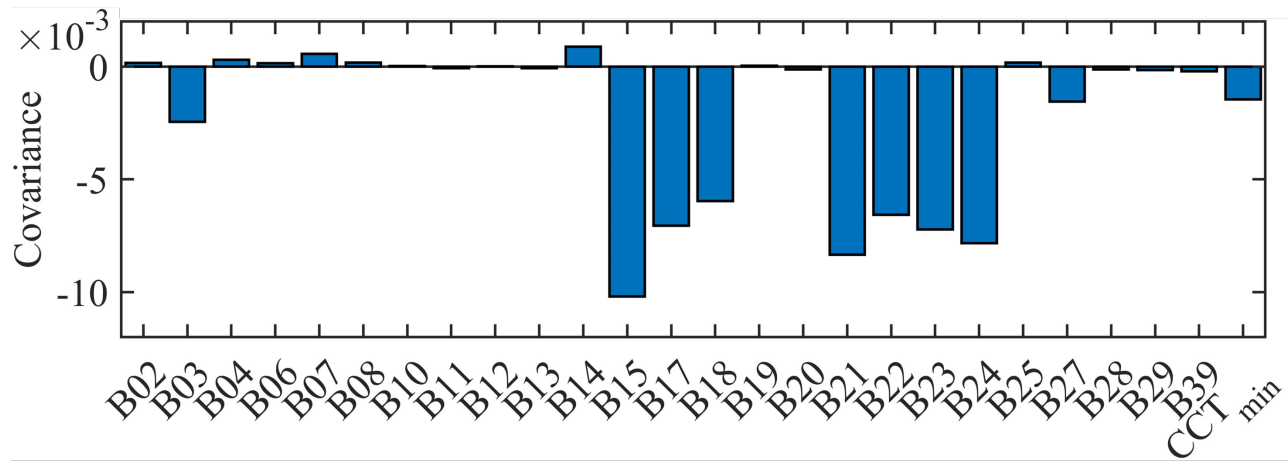

Fig. 12. Covariance between G07 active power loading and G07 active power loading SHAP values at all locations for all n+1 ML models (n=25).

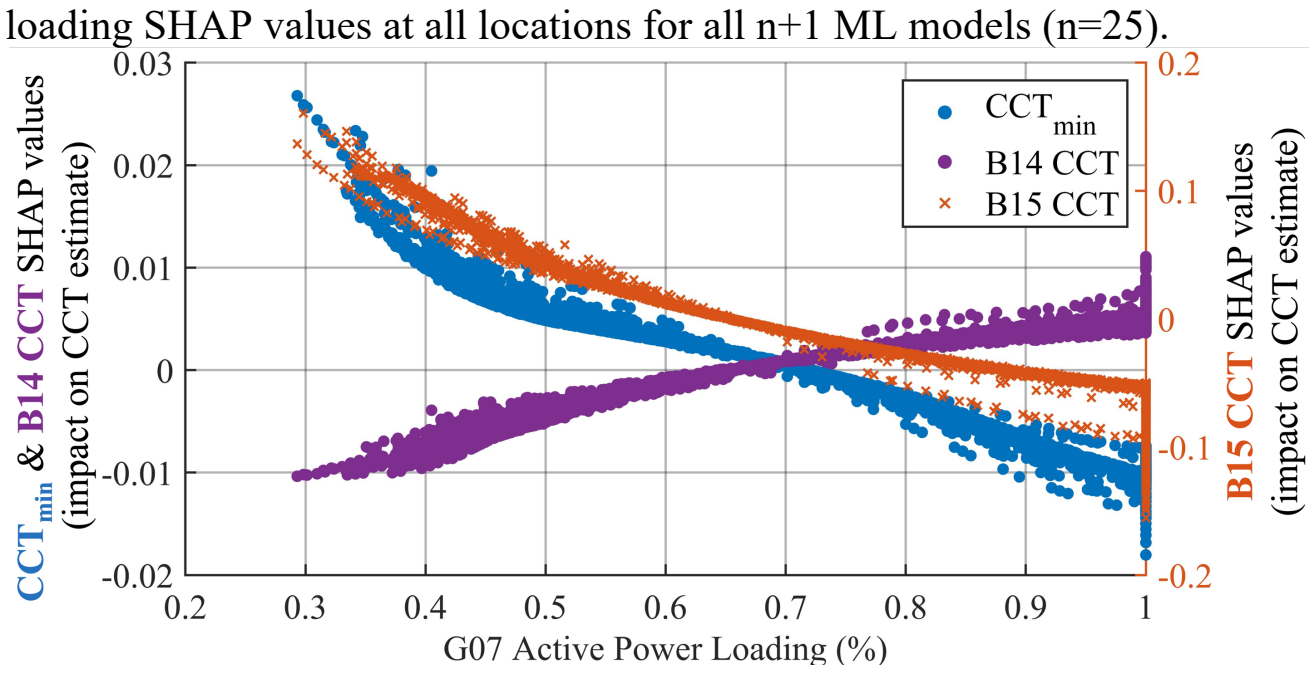

Fig. 13. Impact of G07 active power loading on for $CCT_{min}$, B15 and B14 model SHAP values.

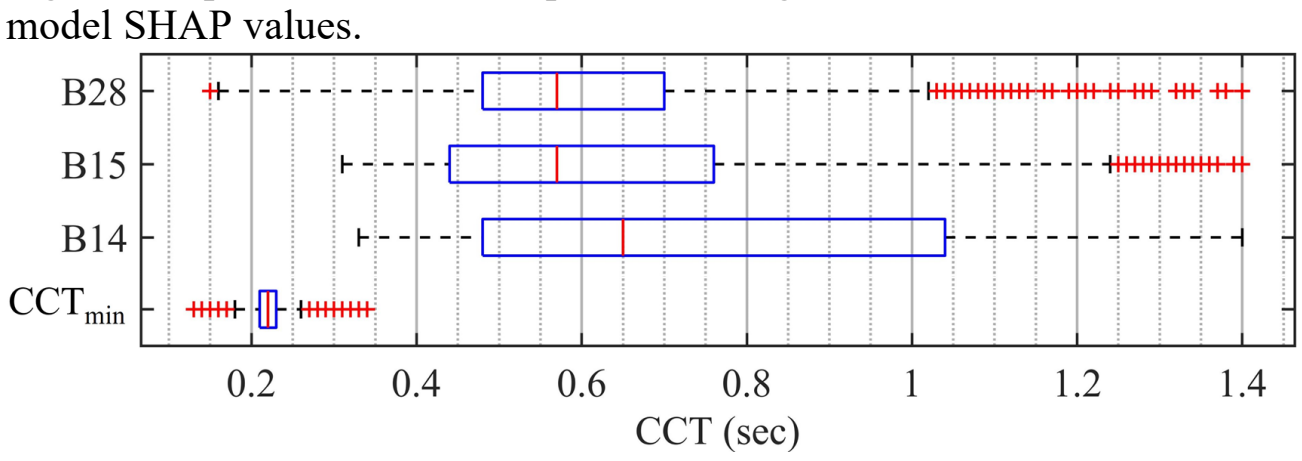

Fig. 14. Actual CCT ranges for key fault locations of interest.

*2) Example B; system-wide impact of RES3 penetration:* Fig. 15 shows the covariance between RES3 percentage of total A3 generation capacity and the stability limit at each location. An increase in the share of A3 generating capacity coming from RES3 has a negative impact (i.e., a reduction) on the CCT at almost every location (including the critical fault, $CCT_{min}$). Fig. 16 shows how B28 CCT is significantly impacted by an increase in RES3 share of A3 generating capacity, with B28 CCT prediction reducing by up to 0.02 sec. In addition, $CCT_{min}$ reduces by up to 0.005 sec. Whilst this is a relatively small value, this is only one factor influencing $CCT_{min}$ (which tends to be very short (Fig. 14).

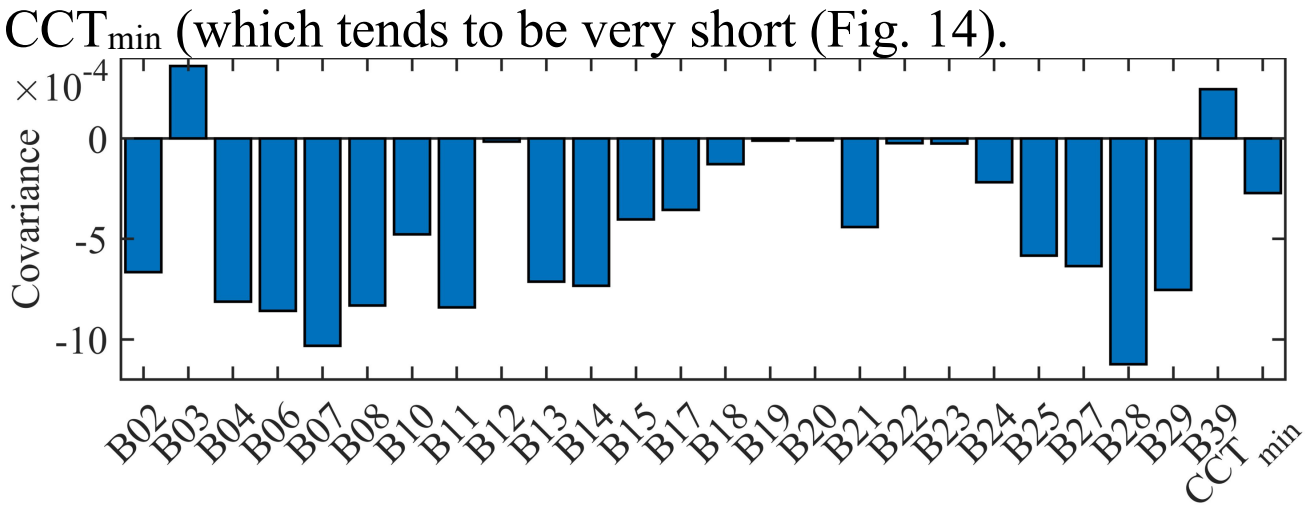

Fig. 15. Covariance between RES3 percentage of total A3 generation capacity and locational SHAP values for all n+1 ML models (n=25).

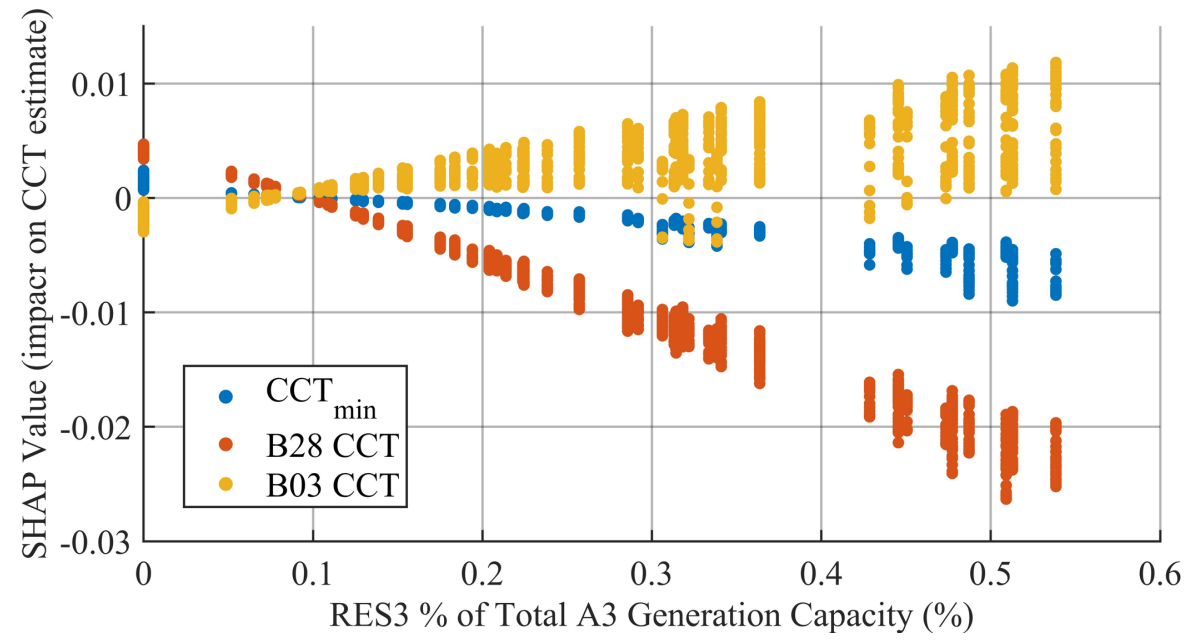

Fig. 16. Impact of RES3 percentage of total A3 generating capacity on $CCT_{min}$, B03 and B28 model SHAP values.

## V. Evaluation of Interventions Based on SHAP Interpretation Rules

This section assesses the impact of a rule derived from a global SHAP interpretation on the CCT at a specific busbar. The important features for $CCT_{min}$ estimation are dominated by active power loading of SGs, therefore B25 is used as an example due to the wide variety of important features – given in Fig. 17. Based on this, a rule could be to: avoid the disconnection of G08 to minimise any reduction of the stability limit at B25. To test the impact of this SHAP-derived operational rule, a new set of RMS-TDS with the rule are run to calculate the CCT at B25. Based on the scenarios defined in Section II. B, there are 423 scenarios where G08 is disconnected. Therefore, an additional 423 RMS-TDS are re-run without G08 disconnection (i.e., the MVA rating remains unchanged, but RES and demand vary as in Section II. B) and B25 CCT calculated. The boxplots of B25 CCTs for these two sets of scenarios are compared in Fig**.** 18.

Results illustrate that by not disconnecting G08, the stability at B25 improves (Fig**.** 18) in alignment with the SHAP interpretation. The lower extreme across all scenarios increases from 0.19 to 0.25 sec (+0.06 sec or 32% improvement), the median improves from 0.39 to 0.42 sec (+0.03 sec or 8% improvement) and the upper extreme from 0.62 to 0.68 sec (0.06 sec or 10% improvement). Note that the most important rule derived by SHAP in this particular case, can also be reinforced through engineering judgment as B25 is the point of connection of G08 to the grid.

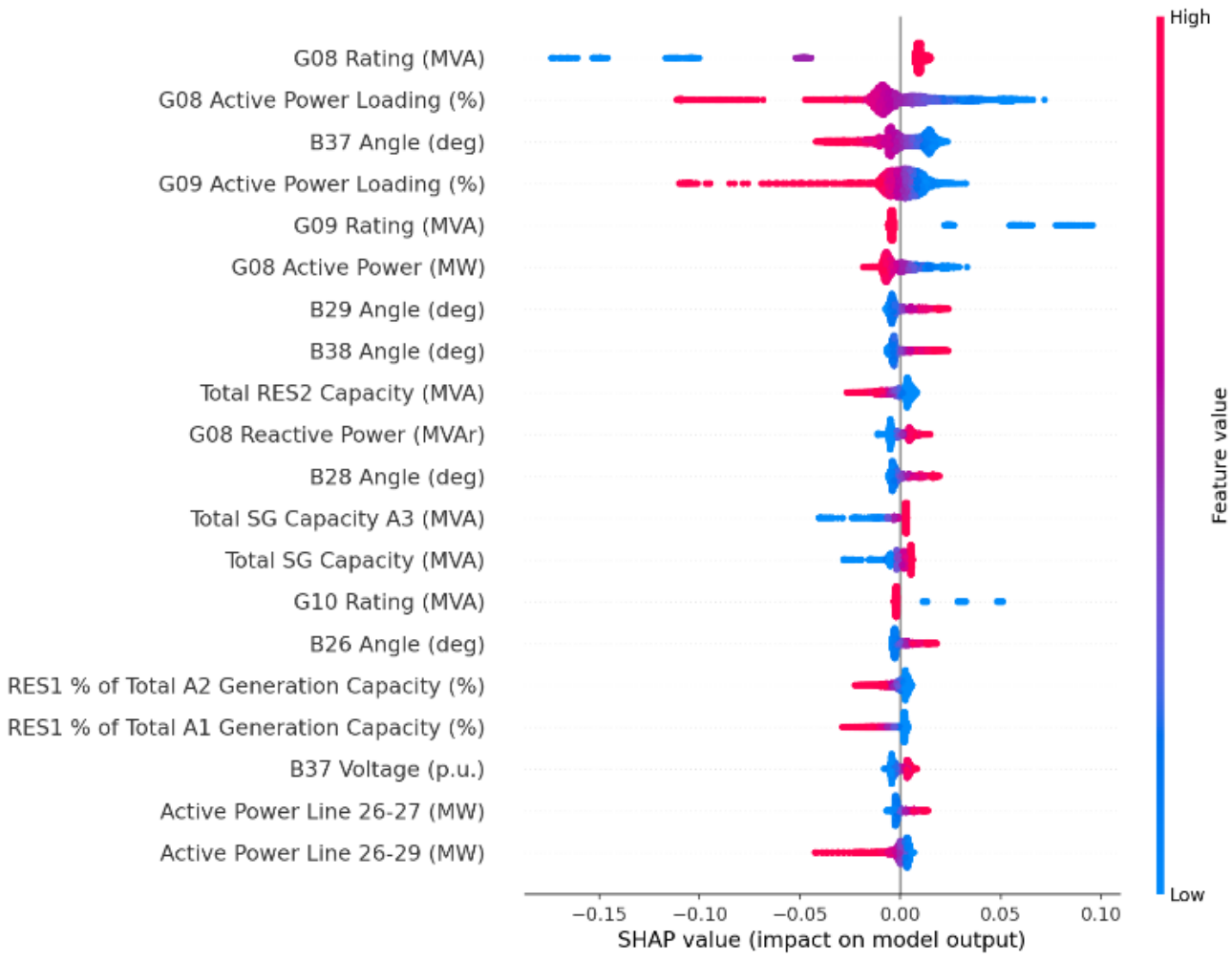

Fig. 17. B25 global SHAP plot.

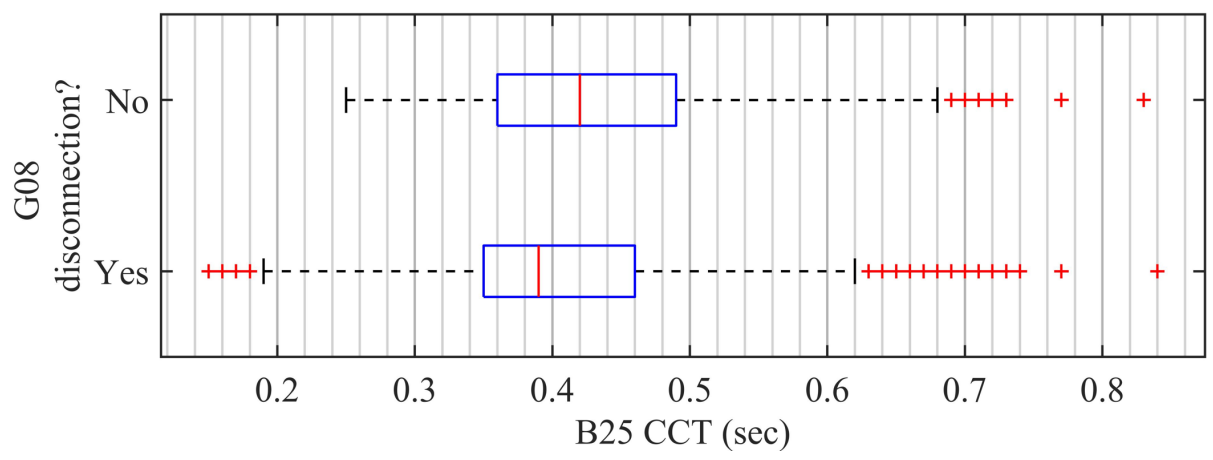

Fig. 18. Improvement from intervention RMS-TDS with no G08 disconnection.

## VI. Conclusions

Transient stability assessment (TSA) in power systems is challenging due to high complexity of system dynamics and high dimensionality – this is exacerbated with the connection of renewable energy resources (RES). In addition, transient stability is location-dependent; meaning that the stability limit may increase at one location and decrease at another. This can lead to spatial changes in the critical contingency, highlighting the importance of methods that take into consideration the effect of contingencies at various locations.

Traditional TSA methods such as time domain simulations (TDS) have limitations with respect to computational requirements, some of which can be overcome using machine learning (ML). However, for ML-based methods to be widely adopted for TSA, the method should be both accurate and interpretable to increase confidence and understanding of underlying prediction mechanisms. Unfortunately, ML algorithms with higher accuracy tend to be opaque, meaning that the user cannot develop an understanding of how the model reaches predictions. This paper proposes a method based on SHapley Additive exPlanations (SHAP) to overcome this challenge [26]. SHAP computes approximate Shapley values, giving the contribution of features in a model to the prediction for a given instance (i.e., local explanation). Local explanations can be extended across all instances, providing a global interpretation of the entire model. Using SHAP for TSA reveals the contribution of power system variables to the prediction of the stability limit of a ML model – in the units used to measure the stability limit (in this instance, sec). The benefit from this can be twofold: (i) to gain increased confidence in ML models and (ii) to reveal trends otherwise difficult to understand due to the increasing complexity and uncertainty of power system operation.

The proposed method involves training a ML model to predict the critical clearing time (CCT) at each busbar $n$ on a network and for different credible operational scenarios. The locational nature of the method is proposed due to the aforementioned locational nature of the transient stability problem itself. An additional model is trained to predict the duration of the critical fault (i.e., the minimum CCT of any busbar for an operational scenario), resulting in a total of $n+1$ ML models. The SHAP framework is subsequently used to gain insights into each of the $n+1$ location-specific models.

The method is applied to a modified version of the IEEE 39 bus networks with RES. We demonstrate the process of ML algorithm selection through trying various ML algorithms until an acceptable level of accuracy is reached – in this case through using artificial neural networks (ANN). We then showcase the types of insights that SHAP can extract form individual ML models (first locally, then globally) and compare the results to permutation feature importance. SHAP is shown to be capable of more detailed insights, by capturing not just feature importance but also feature effects.

This paper also proposes a novel use of SHAP values to identify locational trends in the stability boundary. This is achieved by calculating the covariance between some power system variable of interest (VOI) and the locational SHAP value for that VOI. In doing so, the system-wide impact on the stability limit of changes to that VOI are revealed. The insights provided could be used to formulate rules for system planners and operators, which are only available when using a locational approach as we propose.

## References

[1]J. O'Sullivan, A. Rogers, D. Flynn, P. Smith, A. Mullane, and M. O'Malley, "Studying the maximum instantaneous non-synchronous generation in an island system—Frequency stability challenges in Ireland," *IEEE Transactions on Power Systems*, vol. 29, no. 6, pp. 2943–2951, 2014.

[2]P. Kundur *et al.*, "Definition and classification of power system stability," *IEEE Transactions on Power Systems*, vol. 19, no. 2, pp. 1387–1401, 2004.

[3]R. I. Hamilton, P. N. Papadopoulos, and K. Bell, "An investigation into spatial and temporal aspects of transient stability in power systems with increasing renewable generation," *International Journal of Electrical Power and Energy Systems*, vol. 115, p. 105486, Feb. 2020, doi: 10.1016/j.ijepes.2019.105486.

[4]H. D. Chiang, *Direct methods for stability analysis of electric power systems: theoretical foundation, BCU methodologies, and applications*. John Wiley & Sons, 2011.

[5]B. Iooss and P. Lemaître, "A review on global sensitivity analysis methods," *Uncertainty management in simulation-optimization of complex systems*, pp. 101–122, 2015.

[6]I. A. Hiskens and J. Alseddiqui, "Sensitivity, approximation, and uncertainty in power system dynamic simulation," *IEEE Transactions on Power Systems*, vol. 21, no. 4, pp. 1808–1820, 2006.

[7]K. N. Hasan, R. Preece, and J. v Milanović, "Priority ranking of critical uncertainties affecting small-disturbance stability using sensitivity analysis techniques," *IEEE Transactions on Power Systems*, vol. 32, no. 4, pp. 2629–2639, 2016.

[8]O. A. Alimi, K. Ouahada, and A. M. Abu-Mahfouz, "A review of machine learning approaches to power system security and stability," *IEEE Access*, vol. 8, pp. 113512–113531, 2020.

[9]A. R. Sobbouhi and A. Vahedi, "Transient stability prediction of power system; a review on methods, classification and considerations," *Electric Power Systems Research*, vol. 190, p. 106853, 2021.

[10] S. You *et al.*, "A Review on Artificial Intelligence for Grid Stability Assessment," in *2020 IEEE International Conference on Communications, Control, and Computing Technologies for Smart Grids (SmartGridComm)*, Nov. 2020, pp. 1–6. doi: 10.1109/SmartGridComm47815.2020.9302990.

[11] J. Sohoni and S. K. Joshi, "A Survey on Transient Stability Studies for Electrical Power Systems," in *2018 Clemson University Power Systems Conference (PSC)*, 2018, pp. 1–8. doi: 10.1109/PSC.2018.8664072.

[12] G. Baryannis, S. Dani, and G. Antoniou, "Predicting supply chain risks using machine learning: The trade-off between performance and interpretability," *Future Generation Computer Systems*, vol. 101, pp. 993–1004, 2019.

[13] M. He, J. Zhang, and V. Vittal, "Robust online dynamic security assessment using adaptive ensemble decision-tree learning," *IEEE Transactions on Power Systems*, vol. 28, no. 4, pp. 4089–4098, 2013.

[14] T. Guo and J. v Milanović, "Probabilistic framework for assessing the accuracy of data mining tool for online prediction of transient stability," *IEEE Transactions on Power Systems*, vol. 29, no. 1, pp. 377–385, 2013.
[15] N. Amjady and S. F. Majedi, "Transient stability prediction by a hybrid intelligent system," *IEEE Transactions on Power Systems*, vol. 22, no. 3, pp. 1275–1283, 2007.
[16] F. Hashiesh, H. E. Mostafa, A.-R. Khatib, I. Helal, and M. M. Mansour, "An intelligent wide area synchrophasor based system for predicting and mitigating transient instabilities," *IEEE Trans Smart Grid*, vol. 3, no. 2, pp. 645–652, 2012.
[17] C. Molnar, G. Casalicchio, and B. Bischl, "Interpretable Machine Learning -- A Brief History, State-of-the-Art and Challenges," Oct. 2020, doi: 10.1007/978-3-030-65965-3_28.
[18] R. Machlev *et al.*, "Explainable Artificial Intelligence (XAI) techniques for energy and power systems: Review, challenges and opportunities," *Energy and AI*, vol. 9, p. 100169, 2022, doi: https://doi.org/10.1016/j.egyai.2022.100169.
[19] W. J. Murdoch, C. Singh, K. Kumbier, R. Abbasi-Asl, and B. Yu, "Definitions, methods, and applications in interpretable machine learning," *Proceedings of the National Academy of Sciences*, vol. 116, no. 44, pp. 22071–22080, 2019.
[20] C. Molnar, *Interpretable machine learning*. Lulu.com, 2020. Accessed: Sep. 26, 2022. [Online]. Available: https://christophm.github.io/interpretable-ml-book/
[21] C. Molnar *et al.*, "General pitfalls of model-agnostic interpretation methods for machine learning models," in *International Workshop on Extending Explainable AI Beyond Deep Models and Classifiers*, 2022, pp. 39–68.
[22] M. T. Ribeiro, S. Singh, and C. Guestrin, "'Why Should I Trust You?': Explaining the Predictions of Any Classifier," Feb. 2016, Accessed: Aug. 20, 2021. [Online]. Available: http://arxiv.org/abs/1602.04938
[23] M. Chen, Q. Liu, S. Chen, Y. Liu, C. Zhang, and R. Liu, "XGBoost-Based Algorithm Interpretation and Application on Post-Fault Transient Stability Status Prediction of Power System," *IEEE Access*, vol. 7, pp. 13149–13158, 2019, doi: 10.1109/ACCESS.2019.2893448.
[24] L. Breiman, "Random Forests," *Mach Learn*, vol. 45, no. 1, pp. 5–32, 2001, doi: 10.1023/A:1010933404324.
[25] R. I. Hamilton, P. Papadopoulos, W. Bukhsh, and K. Bell, "Identification of Important Locational, Physical and Economic Dimensions in Power System Transient Stability Margin Estimation," *IEEE Trans Sustain Energy*, pp. 1–1, 2022, doi: 10.1109/tste.2022.3153843.
[26] S. Lundberg and S.-I. Lee, "A Unified Approach to Interpreting Model Predictions," May 2017, Accessed: Aug. 20, 2021. [Online]. Available: https://arxiv.org/abs/1705.07874
[27] L. S. Shapley, "Stochastic Games.," *Proc Natl Acad Sci U S A*, vol. 39, no. 10, pp. 1095–100, Oct. 1953, doi: 10.1073/PNAS.39.10.1095.
[28] J. Kruse, B. Schäfer, and D. Witthaut, "Revealing drivers and risks for power grid frequency stability with explainable AI," *Patterns*, vol. 2, no. 11, p. 100365, 2021.
[29] W. Yi and D. J. Hill, "Topological Stability Analysis of High Renewable Penetrated Systems using Graph Metrics," in *2021 IEEE Madrid PowerTech*, 2021, pp. 1–6.
[30] J. Ren, L. Wang, S. Zhang, Y. Cai, and J. Chen, "Online Critical Unit Detection and Power System Security Control: An Instance-Level Feature Importance Analysis Approach," *Applied Sciences*, vol. 11, no. 12, 2021, doi: 10.3390/app11125460.
[31] N. Hatziargyriou *et al.*, "Stability definitions and characterization of dynamic behavior in systems with high penetration of power electronic interfaced technologies," IEEE, 2020. Accessed: Sep. 26, 2022. [Online]. Available:https://resourcecenter.ieee-pes.org/publications/technical-reports/PES_TP_TR77_PSDP_STABILITY_051320.html
[32] R. D. Zimmerman, C. E. Murillo-Sánchez, and R. J. Thomas, "MATPOWER: Steady-state operations, planning, and analysis tools for power systems research and education," *IEEE Transactions on Power Systems*, vol. 26, no. 1, pp. 12–19, 2010.
[33] "DIgSILENT PowerFactory User Manual," 2019.
[34] M. A. Pai, *Energy function analysis for power system stability*. Kluwer Academic Publishers, 1989.
[35] P. Sorensen, B. Andresen, J. Fortmann, and P. Pourbeik, "Modular structure of wind turbine models in IEC 61400-27-1," 2013, pp. 1–5. doi: 10.1109/PESMG.2013.6672279.
[36] R. Preece, *A probabilistic approach to improving the stability of meshed power networks with embedded HVDC lines*. The University of Manchester (United Kingdom), 2013.
[37] R. Hamilton, "GitHub MATPOWER IEEE 39-bus network case building script," *GitHub*, 2021. https://github.com/RobertIHamilton/MatPower_IEEE39Bus.git (accessed May 07, 2021).
[38] D. Huang, X. Yang, S. Chen, and T. Meng, "Wide-area measurement system-based model-free approach of post-fault rotor angle trajectory prediction for on-line transient instability detection," *IET Generation, Transmission & Distribution*, vol. 12, no. 10, pp. 2425–2435, 2018.
[39] D. Mi, T. Wang, M. Gao, C. Li, and Z. Wang, "Online transient stability margin prediction of power systems with wind farms using ensemble regression trees," *International Transactions on Electrical Energy Systems*, vol. 31, no. 11, p. e13057, 2021.
[40] P. Kundur, N. N. J. Balu, and M. G. M. Lauby, *Power system stability and control*, vol. 7. McGraw-hill New York, 1994. Accessed: Aug. 19, 2019. [Online]. Available: http://www.academia.edu/download/28284657/invitation.pdf

## BIOGRAPHIES

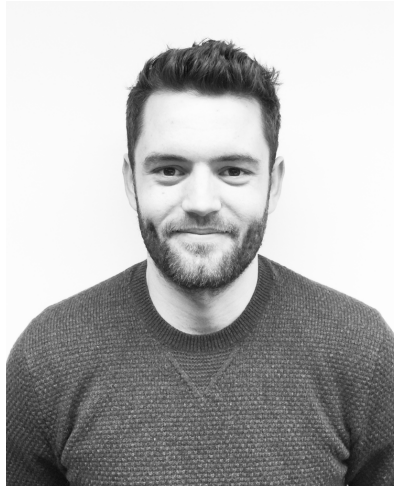

**Robert I. Hamilton** was awarded MEng and PhD degrees from the department of Electrical and Electronic Engineering from the University of Strathclyde in 2016 and 2022. He has since been working as a Research Associate at the university, specializing in transient stability for power systems with high penetrations of renewable generation using interpretable machine learning techniques.

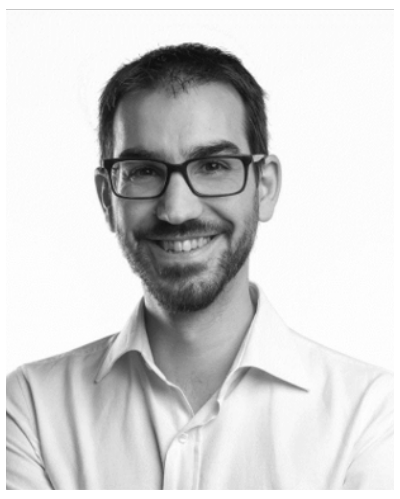

**Panagiotis N. Papadopoulos** (S'05-M'14) received the Dipl. Eng. and Ph.D. degrees from the Department of Electrical and Computer Engineering at the Aristotle University of Thessaloniki, in 2007 and 2014, respectively. Currently, he is a Senior Lecturer and a UKRI Future Leaders Fellow in the Department of Electronic and Electrical Engineering at the University of Strathclyde. His research interests are in power system stability and dynamics, focusing on tackling the increasing uncertainty and complexity in the system dynamic behaviour.